\documentclass[manuscript, screen]{acmart}

\setcopyright{none}
\renewcommand\footnotetextcopyrightpermission[1]{}

\AtBeginDocument{%
  }

\setcopyright{acmlicensed}
\copyrightyear{2025}
\acmYear{2025}
\acmDOI{XXXXXXX.XXXXXXX}

\acmJournal{JACM}
\acmVolume{37}
\acmNumber{4}
\acmArticle{111}
\acmMonth{8}

\usepackage{colortbl}
\usepackage[utf8]{inputenc}
\usepackage{longtable}
\usepackage{array}
\begin{document}

\title{Stakeholder Perspectives on Whether and How Social Robots Can Support Mediation and Advocacy for Higher Education Students with Disabilities}

\author{Alva Markelius}
\authornote{Both authors contributed equally to this research.}
\email{ajkm4@cam.ac.uk}
\orcid{0009-0003-4580-9997}
\affiliation{%
  \institution{Department of Computer Science \& Technology, University of Cambridge}
  \city{University of Cambridge}
  \country{UK}
}

\author{Julie Bailey}
\authornotemark[1]
\affiliation{%
  \institution{Faculty of Education, University of Cambridge}
  \city{University of Cambridge}
  \orcid{0000-0002-3773-7546}
  \country{UK}
}

\author{Jenny L. Gibson}
\affiliation{%
  \institution{Faculty of Education, University of Cambridge}
  \city{University of Cambridge}
  \orcid{0000-0002-6172-6265}
  \country{UK}
}

\author{Hatice Gunes}
\affiliation{%
  \institution{Department of Computer Science \& Technology, University of Cambridge}
  \orcid{0000-0003-2407-3012}
  \city{University of Cambridge}
  \country{UK}
}


\begin{abstract}
This paper presents an iterative, participatory, empirical study that examines the potential of using artificial intelligence, such as social robots and large language models, to support mediation and advocacy for students with disabilities in higher education. Drawing on qualitative data from interviews and focus groups conducted with various stakeholders, including disabled students, disabled student representatives, and disability practitioners at the University of Cambridge, this study reports findings relating to understanding the problem space, ideating robotic support and participatory co-design of advocacy support robots. The findings highlight the potential of these technologies in providing signposting and acting as a sounding board or study companion, while also addressing limitations in empathic understanding, trust, equity, and accessibility. We discuss ethical considerations, including intersectional biases, the double empathy problem, and the implications of deploying social robots in contexts shaped by structural inequalities. Finally, we offer a set of recommendations and suggestions for future research, rethinking the notion of corrective technological interventions to tools that empower and amplify self-advocacy.
\end{abstract}

\ccsdesc[500]{Human-centered computing~Accessibility design and evaluation methods}
\ccsdesc[500]{Social and professional topics~People with disabilities}
\ccsdesc[500]{Human-centered computing~Participatory design}

\keywords{Social Robots, Disability, Mental Health, Higher Education, Human-Robot Interaction, Large Language Models}


\maketitle

\section{Introduction}
In the UK, 20\% of students reported having a disability during the academic year 2022-2023 \cite{disabledstudentsuk2024} and despite existing support, disabled students are still disadvantaged, experiencing a range of barriers to accessing effective adjustments and accommodations \cite{brewer2023disabled}. Existing power dynamics, social injustices and structural barriers may exacerbate challenges related to support and advocacy, limiting some students' ability to articulate their needs effectively \cite{shaw2024inclusion}. This disparity highlights an increasing need for alternative approaches to student advocacy that may empower students with disabilities in ways that current practices may not. While human disability support practitioners can play a crucial role in bridging gaps between students and institutions, these efforts are resource-intensive, relying on trained personnel, availability, and sustained institutional commitment. 

This study explores the feasibility and ethical implications of employing artificial intelligence (AI) and in particular social robots as tools for mediation and advocacy for disabled students in higher education. While the overarching focus regards social robots and LLMs, the study adopts a broader perspective of understanding the use of technology and AI in general for disabled students, to draw insights and identify patterns that can inform the design, implementation, and ethical considerations of AI-driven assistive technologies. Understanding stakeholder perspectives on (AI-based) technologies aimed towards disabled students provides crucial insights into the design and development of LLM-based social robots, ensuring that these systems are adhering to the lived experiences of disabled students. Social robots have previously been used in educational settings, for example as schoolwork companions, coaches for students with Attention-deficit/hyperactivity disorder (ADHD) \cite{O’Connell_Banga_Ayissi_Yaminrafie_Ko_Le_Cislowski_Mataric_2024, lalwani2024productivity}, to reduce anxiety for students with mental health challenges \cite{Rasouli_Ghafurian_Nilsen_Dautenhahn_2024, Matheus_Vázquez_Scassellati_2022} or to enhance skills for students with sensory impairments \cite{Khasawneh_2024}.  Furthermore, large language model (LLM) based AI have been implemented for students with disabilities, for example as conversational accessibility support \cite{Fiora_Piferi_Crovari_Garzotto_2024}, as well as within tools used by disabled students directly in their academic work, including as assistive technology (AT) such as speech to text (e.g. Dragon\footnote{https://www.nuance.com/}, Otter.ai\footnote{https://otter.ai/}), task planning (e.g.  Goblin Tools\footnote{https://goblin.tools/}) and academic writing (e.g. Grammarly\footnote{https://www.grammarly.com/}, ChatGPT\footnote{https://chatgpt.com/}). 

However, existing research involving social robots in educational settings is mainly focused on school contexts, with very limited research in higher education settings. LLMs have been explored as virtual assistants for disability disclosure \cite{lister_taylor_2021} or as facilitators of personalised learning, real-time feedback, content accessibility, adaptive assessment or cognitive support \cite{Neha_Kumar_Sankat_2024}. A few studies have combined social robots and LLMs for disabled students for therapeutic purposes, for example to deliver cognitive behavioural therapy \cite{Kian_Zong}. The potential for social robots to support broader access to learning in higher education, such as self-advocacy for disabled students, is unexplored to the authors' knowledge. 

\subsection{Current Study}
Whereas previous research involving social robots is predominantly focused on academic- or treatment-related interventions for disabled students in higher education \cite{O’Connell_Banga_Ayissi_Yaminrafie_Ko_Le_Cislowski_Mataric_2024, lalwani2024productivity, Khasawneh_2024}, this study instead explores the potential of social robots in supporting self-advocacy in accessing learning in higher education. This study seeks to understand \textit{if} and \textit{how} social robots could help facilitate these needs. Furthermore, the study also directs attention towards LLMs, and in particular social robots combined with LLMs as a foundation for interaction. Given the increased adoption of LLMs as a basis for interaction with social robots \cite{markelius_design_justice, spitale2023vita, Bertacchini_Demarco_Scuro_Pantano_Bilotta_2023} it is crucial to understand the implications of the combination of the two technologies in the context of disabled students. 

Specifically, this paper presents the preliminary findings of a qualitative investigation involving interviews and focus groups with key stakeholders, including students and disability practitioners. The study adopts an iterative participatory and co-design approach, so as to initially understand disabled students' needs and perspectives related to existing advocacy practices. This approach aims to shy away from potentially exploitative and harmful technology (co-)design practices \cite{sloane2022participation, costanza2020design} which relies on the labour and perspective of disabled people to further harmful notions of 'treating the disabled user' with technological interventions \cite{Bennett_Rosner_2019, sloane2022participation}. 

Another key consideration in this work is alignment with a social versus a medical model of disability in technology, AI and robotics research \cite{oliver2013social}. Unfortunately, a substantial amount of social robotics research related to disability is still adhering to a medical model of disability \cite{Frennert_Persson_Skavron_2024}, thus framing disabilities as problems to be 'corrected' through technological interventions, to make individuals conform to normative societal notions of functionality, and thus perpetuating social injustices \cite{Francis_2015, Goering_2015, Bennett_Rosner_2019}. Instead, in this study, we aim to actively view the problem space as the environmental, social and structural barriers disabled students face, instead of positioning the students themselves as the problem. Furthermore, the disabled students' perceptions, self-proclaimed problem space and perspectives are centred as leading variables in the research, shaping not only the findings, but how the social robot development is carried out from a fundamental perspective.

Building on the social model of disability, this study takes a neurodiversity approach in which neurodevelopmental conditions such as autism, ADHD and dyslexia are considered in terms of difference rather than 'disorder' or 'deficit'' \cite{baron2017editorial}. Autistic students make up almost 40 percent of disabled students in UK universities \cite{disabledstudentsuk2024} . Therefore, concepts of empathy in the context of disabled student populations also draw on what is termed the 'double empathy problem'\cite{Milton_2012}. Communication between autistic and non-autistic individuals is prone to mutual misunderstandings \cite{heasman2018perspective}, misinterpretation of mental states \cite{sheppard2016easy}, and negative first impressions \cite{sasson2019first}. Therefore, this study seeks to work with disabled students to understand the communicative aspects of advocacy from within the students' frame of reference. 

The overarching research objective of this study is to \textit{understand whether and how social robots can support mediation and self-advocacy for disabled students}. More specifically, the objectives are related to two main areas of inquiry, with respective research questions. The first phase is related to \textbf{understanding the problem space}, which is crucial to shift the focus from traditional, potentially exploitative development of technologies for disabled people posing technologies as a 'fix' to a normative notion of functionality. By focusing on understanding the existing and changing needs, problems and perspectives of disabled people, this instead allows for an approach which is empowering and inclusive. The research questions for this area are:
\begin{itemize}
    \item     What strategies do disabled students currently employ to support communicating their needs and accessing adjustments?
    \item How do disabled students currently engage with and navigate existing technologies, and how do these technologies shape their self-advocacy and access to learning?
\end{itemize}

The second phase of inquiry is \textbf{envisioning social robots for students with disabilities}, which once again includes students with disabilities to envision the potential role and functionalities of social robots as well as challenges and frictions. Instead of posing a pre-defined role, we wish to enable disabled students to themselves envision if and how a social robot might bring value into their life. The research questions for this area are:
\begin{itemize}
    \item How do disabled students envision and perceive the functional, relational, and affective roles of social robots in supporting mediation and advocacy?
    \item What challenges, ethical concerns, and socio-technical assumptions do disabled students identify in their envisions of social robots, and how do these experiences reflect broader systemic exclusions and inequities?
\end{itemize}

\section{Background and Related Work}
This section presents a background to the context of the study and gives a brief review of previous research relevant to the current study. The first subsection discusses the context of self-advocacy for disabled students in higher education. The second subsection summarises previous work on social robots and LLMs used for people with disabilities in general, and specifically in terms of supporting effective communication of needs, including in educational contexts. 

\subsection{Disability Advocacy in Higher Education}
In UK universities, disabled students have rights under the Equality Act 2010 \cite{uk_equality_act_2010} to reasonable adjustments to prevent disability discrimination. Whilst the rights have not changed since 2010, the interpretation of what is reasonable is constantly evolving in response to changes in understanding of disability and in the educational context. For example, the previous lack of recognition of autism in women and girls has resulted in later diagnosis (often during higher education) for autistic women \cite{lockwood2021barriers}. Updates add to guidance around the interpretation of the Act, such as the recent advice note following the judgment in the legal case of University of Bristol vs Abrahart \cite{EHRC2024}. Such updates act to provide clarity around Higher Education Institution (HEI)'s responsibilities in meeting the needs of disabled students, but take time to fully integrate into practice.  Furthermore, both the number and proportion of disabled students in UK universities have grown significantly over recent years, currently forming over 20\% of UK home students \cite{disabledstudentsuk2024}  therefore, disabled students' self-advocacy operates within a dynamic and evolving context.

In addition to the right to reasonable adjustments, disabled students have access to support to non-medical helpers (NMHs), assistive technology and funding to meet other access needs (transport, printing costs, etc.) determined through a needs assessment \cite{govuk_dsa}. Therefore, a disabled student is likely to be navigating a combination of commonly used adjustments (such as additional time in examinations) and more individualised adjustments.  Within each HEI, the disability service has a key role in advising on what is (and is not) reasonable in terms of adjustments.  Structural factors and power dynamics in higher education have long been recognised to marginalise disadvantaged groups within higher education \cite{heffernan2022bourdieu}. Tensions are inherent in the structure of support for disabled students when access to information and guidance are controlled and funded by the institution itself. Therefore, a further key dimension of support for disabled students is through student-led groups and independent advice (e.g. university student union, disabled students campaigns, etc.) who often provide a source of advice and support independent of the HEI. 

Self-advocacy for disabled students has been defined variously, from narrow definitions around disclosure of disability in order to access adjustment to a much broader and more complex set of ideas and behaviours around the individual's understanding of themselves and their relationship with the individuals and structures around them \cite{smith2022importance}. Importantly, self-advocacy involves a level of assertiveness in communication (written, spoken or body language) that requires knowledge of disability rights, organisational structures and resources, and self-knowledge \cite{test2005conceptual}. Explorations of the self-advocacy experiences of disabled students reveal the relational complexities inherent in negotiating institutional process and relational complexities \cite{bruce2021disability}. 

The example of lecture capture illustrates this dynamic and evolving landscape which disabled students  self-advocate to access their learning. Access to recordings of lectures is commonly recommended as an adjustment that allows the disabled student flexibility to manage their studies. Studies of the use of lecture capture have shown that this is an important strategy for disabled students \cite{horlin2024can}. However, the 'reasonableness' of lecture capture is fiercely debated within institutions, creating tensions, complexity and inconsistency in access to this adjustment \cite{nordmann2022lecture}. Students therefore frequently find themselves in the position of having to ask (sometimes repeatedly) for access to lecture recordings even after they are identified as an appropriate adjustment on student support documentation \cite{horlin2024can}. 

Specific considerations for advocacy can be anticipated for disabled students with communication preferences. For example, autistic individuals report a strong preference for non-verbal communication, preferring the clarity of written communication \cite{howard2021anything}.  Challenges in navigating higher education can be complex and cumulative, with overwhelm, frustration and exhaustion commonly reported by neurodivergent students \cite{bailey2024neurodiversity}.  Therefore, in addition to achieving a  successful outcome in terms of accessing specific adjustments, disabled students are balancing overall capacity and energy levels \cite{brewer2023disabled}, with a desire to effect meaningful change around disability discrimination \cite{smith2022importance}. 

\subsection{Social Robotics, LLMs \& Disability}
There exists to date a noteworthy body of research aimed at developing social robots, LLMs and other forms of AI in broader disability contexts \cite{markscoping} ranging from navigation robots for blind people, \cite{guerreiro_cabot_2019}, social robots supporting learning for people with intellectual disability \cite{mitchell_social_2021} or assistance for people with executive dysfunction \cite{dam_experiences_2022} to avatar robots enabling telework for people with mobility disabilities \cite{Takeuchi_Yamazaki_Yoshifuji_2020} and virtual reality chatbots supporting people with intellectual disabilities in every day life tasks \cite{Garcia-Pi_Chaudhury_Versaw_Back_Kwon_Kicklighter_Taele_Seo_2023}. A recent scoping review \cite{markscoping} found that LLMs and social robots have been implemented for a range of different disabilities, including neurodivergence, cognitive impairment, mental health, mobility impairment, sensory impairment and chronic health conditions. They report that a majority of studies from the last decade have focused on HRI only, but that an increasing trend is the usage of LLM-based social agents as well \cite{lister_taylor_2021, A_V_Jebadurai_Vedamanickam_Kumar_2023, Garcia-Pi_Chaudhury_Versaw_Back_Kwon_Kicklighter_Taele_Seo_2023}. A small number of studies have combined the two technologies for interventions with disabled people, including a voice controlled feeding assisting robot for people with mobility impairment \cite{Padmanabha_Yuan_Gupta_Karachiwalla_Majidi_Admoni_Erickson_2024} and a social robot connected with ChatGPT to support and improve cognitive
functioning for autistic people \cite{Bertacchini_Demarco_Scuro_Pantano_Bilotta_2023}. Still, there exists several gaps in the research, and we highlight in particular i) the need to understand social robots and LLMs \textit{combined} social, ethical and functional impact for students with disabilities and ii) the need for research of social robots and LLMs as assistive technologies to acknowledge, discuss and engage with models of disability as discussed in section 2.2.2.

\subsubsection{Previous Work with Disabled Students}
The scope of work relating to using social robots and LLMs for disabled students is very broad, but the total body of literature remains small \cite{Padmanabha_Yuan_Gupta_Karachiwalla_Majidi_Admoni_Erickson_2024}. Furthermore, a smaller number of studies are aimed specifically at disabled students in higher education. One example, however, is Taylor \cite{lister_taylor_2021}, an LLM-based virtual assistant for disability disclosure, developed through participatory design with disabled students. The chatbot is implemented to ease and mitigate administrative burden in higher education, such as form-filling and bureaucracy, which often poses a significant barrier and is seen as a serious issue. Educational robots have also been investigated to assist developing programming skills among students with hearing disabilities \cite{Khasawneh_2024} i.e. through collaborative learning to develop programming and related skills as well as developing motivation toward learning programming. Furthermore, a socially assistive robot has been developed and implemented as an in-dorm study companion for college students with ADHD \cite{O’Connell_Banga_Ayissi_Yaminrafie_Ko_Le_Cislowski_Mataric_2024}. The authors of this study also adopted a user-centred and participatory design over a three-week intervention. Students reported positive feedback after evaluating usability and functionality of having the robot present during study sessions e.g. through a sense of companionship and accountability. Similarly, a robot called Productivity CoachBot \cite{lalwani2024productivity} was developed as a coach for university students with ADHD, designed to engage in conversations, generate daily schedules, send task reminders, and monitor structured study sessions. The social robot is connected to ChatGPT API, and showed promising results in an early evaluation without clinical subjects. Finally, a number of studies are addressing mental health related disabilities. One study combined LLMs and a socially assistive robot to deliver cognitive behavioural therapy (CBT) to university students \cite{Kian_Zong}. The study reported that self-reported measures of general psychological distress showed a significant decrease over the study period. These findings indicate that SAR-guided, LLM-powered CBT may be as effective as traditional worksheet-based methods in facilitating therapeutic progress throughout the study and may surpass them in reducing user anxiety immediately following a CBT exercise. Another intervention developed a social robot that supports deep breathing practices for the purpose of reducing anxiety, by placing the user's hands on a robot that expands and contracts \cite{Matheus_Vázquez_Scassellati_2022}. The robot evaluation showed a significant measurable reduction in anxiety state among the participants. Not only have social robots been investigated in relation to mental health among university students, but also intelligent agents in general and their potential to support students coping with stress and anxiety \cite{rasouli2024university}. A recent survey \cite{rasouli2024university} observes that students value adaptability, confidentiality, and accessibility for different types of intelligent agents, e.g. virtual agents and social robots, and those with anxiety were willing to use agents for activities to increase well-being. The studies above show that integration of LLM-based social robots for students in higher education, including those with disabilities, is already underway. However, the majority of these studies relate to addressing specific, pre-determined issues, without holistically investigating from the disabled students perspectives what potential support robots may offer. These studies often focus on predefined challenges, such as assistive functionalities, rather than exploring the broader role that social robots could play in enabling agency, autonomy, and meaningful participation for disabled students. Moreover, research tends to adopt a top-down, solution-oriented approach, where the perspectives of disabled students are considered reactively rather than proactively shaping the design and implementation of these technologies. This risks reinforcing normative assumptions about disability and overlooking the lived experiences shaping situated needs and preferences of disabled students. To address these gaps, this research aims to engage directly with disabled students in co-creative, participatory processes to critically assess how (LLM-based) social robots might support—not just compensate for—different aspects of their academic and social experiences.

\subsubsection{Rethinking Social Robots for Disabled People}
Previous work has not only developed and implemented interventions of social AI agents for disabled people, but scholars have also engaged in inquiry about the ethical, functional and systematic challenges inherent in this development, and how to rethink it to avoid potentially harmful and exploitative practices. Despite the potential benefits from using both LLM-based virtual social agents and social robots for disabled people, several studies raise concerns about biases \cite{Urbina_Vu_Nguyen_2024}, dehumanising tendencies \cite{Nakamura_2019}, marginalising trends such as working within a medical model of disability \cite{Frennert_Persson_Skavron_2024} and ontological concerns such as the double empathy problem \cite{Milton_2012}. Previous research shows that LLMs such as ChatGPT and Gemini\footnote{https://gemini.google.com/} are prone to ability bias \cite{Urbina_Vu_Nguyen_2024} through significant underestimation of capabilities, and framing of fewer favourable qualities and more limitations for disabled people. Indeed, a comprehensive study aimed at identifying and quantifying bias against stigmatised groups in LLMs found evidence of high probability of negative attitude related to physical traits, diseases and disability \cite{Mei_Fereidooni_Caliskan_2023}. Furthermore, an empirical study conducted 19 focus groups with 56 participants from the disability community to identify and categorize potential harms perpetuated by LLMs toward individuals with disabilities \cite{Gadiraju_Kane_Dev_Taylor_Wang_Denton_Brewer_2023} and found e.g. validating incorrect perceptions and teaching dangerous ways of interacting with people with disabilities as significant prevalent harms. 

But issues like this are not only found in LLMs; practices among social robotics applications for disability have also been subject to warranted critique \cite{rizvi2024robots, Nakamura_2019, williams2021misfit}. For instance, scholars drawing upon Critical Disability Studies \cite{meekosha2009s} and Crip Theory \cite{mcruer2008crip} have highlighted how applied robotics in autism intervention often perpetuates problematic narratives. Roboticists frequently position care work as a legitimising application for their creations, however these interventions often draw on dominant, harmful narrative tropes of autism \cite{williams2021misfit}. Indeed, \cite{dehnert2024ability} highlights how social robotics often replicates societal biases, reinforcing norms of the “socially constructed normal body”. For instance, neurotypicality frequently serves as the implicit standard for social robots, dictating their design and behaviour while marginalising neurodivergent ways of interacting and communicating. This perpetuates exclusionary practices and overlooks the agency of disabled individuals in shaping technology. The author calls for a deeper engagement with disability justice principles in the field of social robotics. This means prioritising the lived experiences of disabled people at every stage of design and research. It also requires acknowledging neuronormativity \cite{saetra2022normativity}, intersectionality and power dynamics particularly in how we understand and define "ability" and "disability" within human-robot interactions. Furthermore, the combination of social robotics and LLMs remain under-explored \cite{markelius_design_justice}, and so also the social and ethical implications of their combination. Thus, to fully understand how combining the two technologies might impact the communities for which they are intended, research has to be aimed to understand their social and ethical implications individually and combined to be able to address them. This is a research gap this study also aims to take a first step towards investigating.

\section{Research Context}
This study adopted an iterative and participatory design approach, mainly relying on qualitative data collection from focus groups and interviews. The approach was split into two phases, an initial interview phase (Phase 1), followed by a focus group phase (Phase 2) which was informed by the interviews. Informed consent was obtained from all participants and a topic guide was provided before the sessions to enable familiarisation with the relevant themes and glossary of terms as seen in Appendix A.1-4. The study received ethical approval from the Department of Computer Science \& Technology, University of Cambridge (Ref. Ethics Review \#2351: How Technology, AI and Robots Can Help for Mediation and Advocacy for Students with Disabilities). 

The research was carried out at the University of Cambridge, with all participants either current students or current experienced disability practitioners. Disability support at the University of Cambridge is led by the Accessibility and Disability Resource Centre (ADRC) which sits within the central university's Student Support Services. Students are also members of a college, which currently do not offer specific disability support, but do support students wellbeing and often directly provide accommodation, study spaces and catering, from which accessibility issues can arise. Furthermore, colleges frequently play a role in supporting students in accessing adjustments and intervening when issues arise.  Students registered with the ADRC are allocated a Disability Advisor and have a student support document (SSD) which serves as a guide for adjustments and is shared at the students' request with relevant staff across the collegiate university.  The ADRC directly employs non-medical helpers (NMHs) to take on the support roles identified in the SSD.  A key NMH role is that of the disability mentor who works one-to-one with students over the course of their studies offering advice, support and guidance on all aspects of accessing life at university. It is not uncommon for the student-mentor relationship to continue across several years, particularly if the student goes on to complete further degrees at the university.  Whilst disability mentors are employed by the university, they have very limited direct contact with other organisations or individuals, acting instead as a quasi-independent resource and ally for the student. 

In line with national trends \cite{di2024diversity},  the number of students disclosing a disability has risen considerably in recent years, from 1,200 in 2010 \cite{DRC2010}  to over 6,300 in 2024 \cite{ADRC2025}, translating into over 22,000 hours of non-medical help delivered in the academic year 2023-24.  Although student survey data reports 93\% student satisfaction with the provision \cite{ADRC2025}, some student face challenges in accessing appropriate provision.  Although only a relatively small number of students' experiences (88 responses) from the University of Cambridge were captured in the recent Disabled Students UK 2024 Access Insights Report, the proportion of disabled students who felt that they have the support they need was the lowest at 26\% of the UK universities included in the report, and the highest proportion of students who had been made to feel unwelcome by a member of staff due to their disability \cite{disabledstudentsuk2024}. Furthermore, the report highlights issues with access to information about different adjustments and the effective provision of agreed adjustments.  

In exploring the potential for social robots to complement and support existing provision within our own institution, we presents a case study of the University of Cambridge.  As such, the study reflects features of disability student support that are specific to the context of this institution, reflecting its unique strengths and challenges. However, many of the features explored, and challenges presented, apply more broader to the higher education context in the and beyond. 

\section{Phase 1: Interviews}
Interviews were conducted during the first term of the 2024-25 academic year with four stakeholders, all current students or disability practitioners at the University of Cambridge. The interviews were designed to support the interviewee in expressing and exploring their experiences and perspectives as freely as possible within their own frames of reference. Furthermore, care was taken in the data collection and analysis to seek understanding beyond a superficial 'barriers and enablers' approach to a deeper understanding of the complexity of the relationship between present and potential forms of advocacy support for disabled students \cite{haynes2024rethinking}.

\subsection{Participants}
Interviews were carried out with four participants (2 experienced disability practitioners and 2 disabled students). During  the interview, both of the disability practitioners also disclosed disability.  Practitioner 1 (P1) is a disability advisor with over 20 years experience of supporting disabled students. Practitioner 2 (P2) is a disability mentor (alongside other roles) with 30 years of experience in working with disabled students across a range of contexts. Student 1 (S1) is a disabled undergraduate student. Student 2 (S2) is a disabled undergraduate student who is also a Disability Representative at their college. Both students have declared disability conditions and are registered with the ADRC (Accessibility and Disability Resource Centre; the student disability support organisation for the University of Cambridge). Participants' disabilities include a range of conditions. The small sample size and discussion of other aspects of the participants' roles and contexts prohibits further details of their nature of their disability due to the risk to anonymity. Participants were compensated with an Etsy or Amazon voucher of £12.   
 
\subsection{Procedure}
Interviews lasting one hour were carried out individually at a time to suit the participant at a location convenient for the participant within the University of Cambridge. The interviews were recorded and live transcribed within MS Teams, with the transcripts reviewed and edited for accuracy using the audio recordings.  A detailed Topic Guide was provided in advance of the interview in order to (1) ensure that participants were aware of  and comfortable with, the scope and foci of the interview, and (2) were familiar with terminology. In addition, a brief video was used to define social robots.  The foci of the interview were: (1) Advocacy and communication needs, (2) The use of AI-based tools , (3) The use of social robots, (4) The use of robots in student support, and (5) Ethical and practical concerns around robots and AI. A detailed overview of topic guides for the interviews can be found in Appendix A.2. Questions were kept as open and brief as possible, with follow up questions and prompts supporting the participants in exploring their experiences and viewpoints more fully within their own frames of reference. A glossary of terms was provided for all participants as seen in Appendix A.1. 

\subsection{Data Analysis}
The interview data were analysed following the procedure for reflexive thematic analysis \cite{braun2006using}. The development of themes focused on allowing the underlying dimensional issues around the use and potential of robots to support the self-advocacy of disabled students, to reflect the inherent complexity and move beyond barriers and enablers to a deeper understanding of the issues \cite{haynes2024rethinking}.  

\subsection{Results Phase 1}
\begin{figure}
    \centering
    \includegraphics[width=1\linewidth]{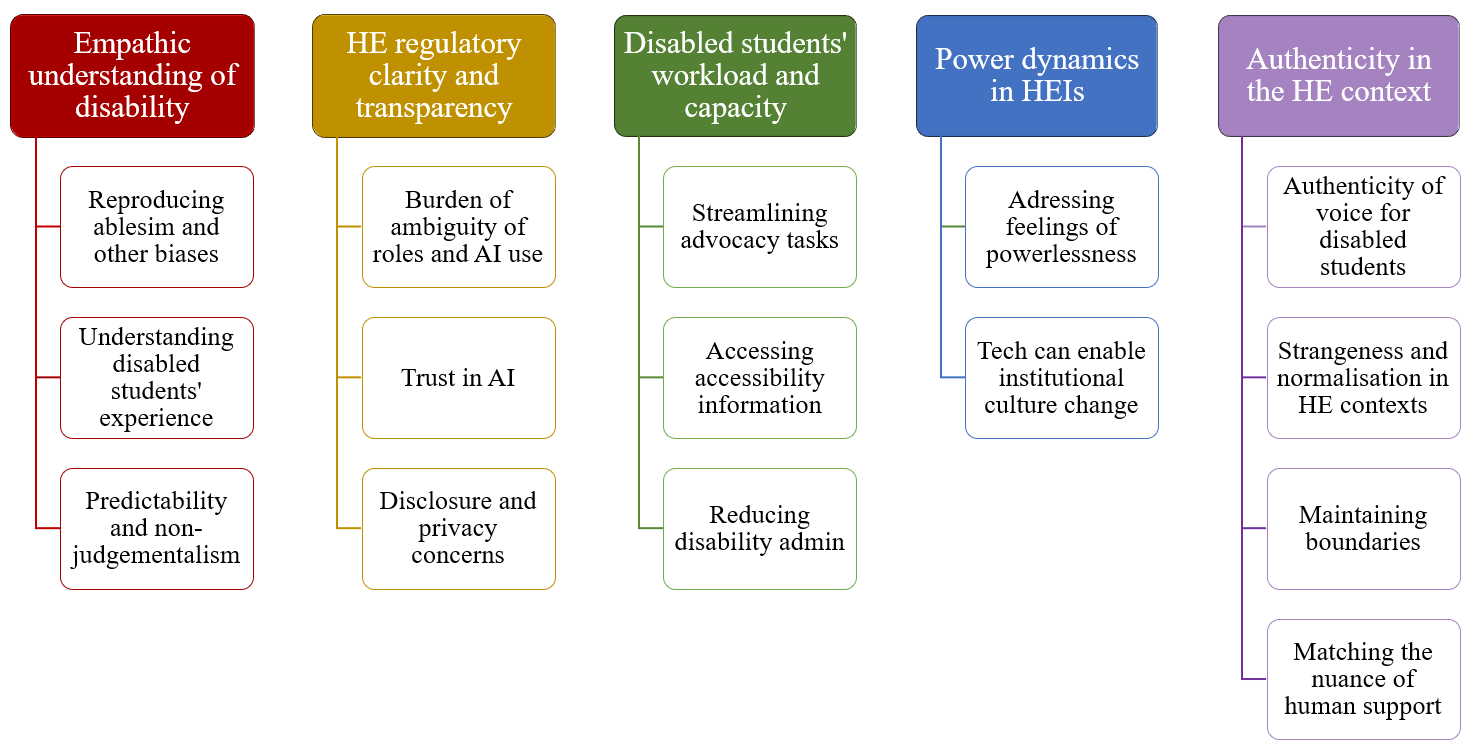}
    \caption{Interview themes and subthemes from thematic analysis of Phase 1 of the study}
    \label{theme}
\end{figure}

Thematic analysis of the interview data resulted in the emergence of five themes, reflecting dimensions of a highly complex, contextually dependent, and dynamic landscape of experiences and perspectives on existing and potential technological tools to support self-advocacy. The themes transcend simplistic positive-negative or barrier-enabler dichotomies. In fact, contradictions and tensions between these binaries are frequent within the themes \cite{haynes2024rethinking}. Instead, these themes reflect a deeper exploration of the dimensions of importance to disabled students and student support workers in engaging with emerging robot technology within an already dynamic and complex higher education disability support context. An overview of the themes for the thematic analysis of the interviews can be seen in Figure \ref{theme}.

\subsubsection{Empathic Understanding of Disability}
Issues and complexity around empathic understanding within disability support were a key focus, both in the context of current experiences of accessing learning and in considering the future potential of robot-aided advocacy. For example,  'how would a robot know that a small gesture or word that could indicate something really huge that actually would be the core of what the conversation needs to be about?' (P2).  S2 felt this was central to providing support for disabled students, as 'there's a massive role for [...] empathy and understanding, especially as it's not easy to be disabled and it's not easy to have that identity... empathy is very, very important in speaking about disability ..., especially if you're a non disabled person talking to disabled person ... when you're talking to people about their disability and what accommodations or adjustments they might need ... their feelings around it can be quite complicated and it can be quite complex to navigate'.

There was particular concern about empathy in the context of disability and diversity, with potential for causing distress as  '[what] might be true for one student but not for everybody. [If the] robot responds inappropriately and they might [think] I don't understand why they're laughing..that's not funny' (P2). 

When considering the potential for a robot helper to observe and interpret expressions and body language, participants considered carefully how this might work in the context of neurodiversity, with S1 reflecting that 'my first thought is would be ... patronising and infantilising, but then ... when I reflect, I think it probably would be helpful because so often [these are] barriers to study for me, I just don't realise that I'm …tired or hungry or just something like that which stops me from doing anything, and it's just the most basic thing, isn't it really?'.  

However, S2 took 'issue with the premise that ... you're always able to identify what someone is in your feeling from ... their non-verbal cues ... Is this robot going to be very rude to autistic people because it can't read their [body language]? ... I can imagine this being trained on a neurotypical population and being like inappropriately applied... to a disabled population'.

Both students and support staff were concerned about \textit{ableism and other societal biases} being reproduced in  technologies trained on existing forms of language and disability understanding, as 'AI is built on what is already existing and what is already existing is quite problematic' (P2). Importantly, intersectionalities, such as 'racism, homophobia… can really compound the experience of discrimination' (S2), as 'AI is built on what is already existing and what is already existing is quite problematic' (S2).  'Biases exist [so] …  I would worry that anything that is meant to help advocate for you or communicate your needs would not guarantee … your safety or that you will be represented in a way that's going to help navigate [a sometimes] very traumatic process' (P2).
 
Participants discussed the challenges in developing a full \textit{understanding of the disabled experience of higher education}, sharing examples of a lack of understanding in the staff implementing adjustments required to support students' access to learning.  P1 explained how 'the social model of disability changed my experience of who I was and how I thought about myself entirely', and P2 emphasising the importance of 'disability as a … political identity is a … useful tool for people because… rather than …internalising problems … it helps … to look at areas where you are held back in a world or an environment which is not set up for you'. 

Understanding the diversity of the disabled student experience through listening and relationship building was mentioned as a key feature of successful student support, with P2 explaining that 'every student has a different pathway [so the] process [of providing support and advice] is quite intuitive and very particular'.  'When you speak to someone, it can be quite useful if they have some sort of knowledge about your condition, so I would worry with a robot that it would based on whatever information set it's ... used. It would maybe think rigidly about your condition and you know not be able to appreciate the diversity of individual people [that] have on paper, the same diagnosis or same condition' (S2). 

When considering potential interaction with a robot, participants identified \textit{predictability and non-judgementalism} as key features that they would welcome. These features were considered particularly important for autistic students. Participants made contrasts between these features and the more challenging interactions they and others had experienced.  P1 explained that 'human beings are too unpredictable, they're too judgmental. There's so much nervousness around being around another human being because [students] just don't know how they're going to react'. P2 explained that 'some people feel that they're talking to a robot, they’ll actually divulge more because they feel it's not going to judge them'. 

\subsubsection{HE Regulatory Clarity and Transparency }
Participants gave examples of the existing \textit{burden of navigating the ambiguity} around the use of AI in higher education, expressing concerns around further effortful management of increasingly integrated and subtle forms of AI. P1 explained that 'I feel it's having an extra burden to students who are using those tools because they are time poor in the first place'.  P1 also expressed concern over the additional burden on autistic students who are more likely to 'take that instruction [to declare AI input in work] literally' and attempt to record every AI input, recalling an autistic student feeling 'really worried he was ... using it behind the scenes for perfectly legitimate reasons, [but felt that] if I hand in my PhD and somebody finds out I've used ChatGPT to convert a document from one form to another, I’ll get hauled up'.  S1 echoed this concern when using AI tools, it feels like 'I .. feel as if I should …cover my screen as if I'm doing something … naughty'. 

Concern over the lack of clarity around the use of AI was expressed, with P1 worried about the adverse impact on disabled students as 'banning [AI tools] outright is profoundly dangerous and has a disproportionate impact on the disabled students using them'.  Participants welcomed emerging clarity around AI usage, with S1 commenting that an AI tool was 'recommended to us … then I thought, oh, this is actually legitimate and allowed. And that's OK because I had been a bit nervous about using AI before ... Sometimes I get in my head … what if I've accidentally taken something from AI] could this be constituted as plagiarism? So, I do try to be really cautious'. 

Issues around the use of AI based technology were frequently framed in terms of \textit{trust}. P2 explaining that 'software rests on quite a thin layer of trust ... If it gets it wrong once even if later after, you know, it comes back and sort of corrects itself, they've lost [trust] something is broken already'.

A core concern for disabled students is navigation of \textit{disability disclosure}, with robot interactions adding uncertainty and complexity,  but potential for more comfortable disclosure, with S1 explaining that 'personally I've gotten to the point where I'm … comfortable asking people for help and … disclosing …, but for a long time that was a big barrier. .. I think if there had been a machine that I could communicate that to you, that would have been a lot easier'. 

However, disabled students and professionals all expressed concerns over data privacy,, not knowing 'what the robot was doing with the information, you know, where are those recordings going. If I'm talking, you know if I'm delving into my past history or I'm talking about the concerns about this, I'm happy to have a robot reflect back to me. But what is going on with all that data?' (P1). S1 explaining that 'I would have a distrust of anything that I knew was taking my data, like processing it and then formulating a response. So I guess I’d want to know who owns the company and where that could end up?'.

\subsubsection{Disabled Students' Workload and Capacity}
The existing and potential role of technology in \textit{boosting the capacity} of students were discussed, both in terms of streamlining academic tasks and in reducing disability admin.  Tasks around advocating for one's needs largely involved either communicating those needs (often repeatedly) or searching for the right information to support effective requests. S2 described 'spend[ing] a lot [of time] emailing my disability advisor with …  tasks that do need to be done'. Tools that help with tasks when 'you're low energy and you just need to get something done' (S2) were particularly valued  by participants. 

The importance of clarity of communication emerged as a key dimension in the experiences and perspectives of disabled students.  Frequently mentioned were the challenges faced by students in clearly communicating their needs,  particularly when student support documentation had been shared but not implemented.  Participants described the frequent use of AI based tools to support advocacy related communications, making emails 'clearer or more succinct' (P2), ensuring 'a positive and respectful tone [and] adding the frills' (S1).  Within the context of the University of Cambridge, an individual disabled student's support may involve the central university services, their department and their college, meaning that 'finding the right person and that can be a bit of a challenge' (P1). 

Of particular interest to disabled students and support staff was the potential for a robot to provide an  interface for \textit{accessing disability-related information}. However, Participants had experience of investing time and energy in engaging with assistive technology that didn't reduce overall workload. S2 was concerned when 'the learning curve is … too slow and the amount of time that you have to invest in learning to use it, or really irritatingly, if you kind of get to a certain point using it and find that it can't actually do what you want it to do, it's just too clunky or it's just too difficult. That can be very frustrating'.

One frustrating aspects of \textit{disability admin} is 'when things are communicated but not actually put into place by the department … I’m meant to have access to lecture recordings and that's just not provided … it's something I've followed up, but it's clear that they're not going to [provide in] my time here … [it’s] really frustrating' (S2). P2 explains that 'advocacy admin is exhausting and AI or a social Robot could be useful for stepping in and sending the email for the student instead of them having to repeat their situations and needs over and again. For repeated asks and tasks AI could be extremely useful', with S2 wondering how much time and effort could be saved 'if you had a robot which could do all that [chasing up adjustments] work for you. 

\subsubsection{Democratising Power Dynamics in HE}
Participants discussed the need for institutional change around disability, often connected to feelings of \textit{powerlessness}. The power imbalance between student and staff member was felt to create barriers to listening and understanding, with P1 explaining in terms of a  'power relationship - if he's gonna be understood … has somebody already made up their mind … that this student is a problem and it's their disability that is the issue rather than listening to what the student is actually saying?'. 'The student had already tried to [put adjustments in place], but wasn't being listened to by the people he could actually action it' (P1). In contrast, 'the times that advocacy support needs goes well is when you can say what you need once [to] someone who listens carefully and then everything just happens' (P2). 

Participants described the frustration and anger that comes from the feelings of powerlessness when students' 'very reasonable adjustments are not being met or being framed as awkward or unnecessary' (P2). 'When things are communicated but not actually put into place by the department … I’m meant to have access to lecture recordings and that's just not provided … it's something I've followed up, but it's clear that they're not going to [provide in] my time here … [it’s] really frustrating' (S2).

Participants felt that tech has a role in reducing the number of promises without action,  when students get 'emails back, saying, oh, don't worry, we'll sort it, but it's never ... quite sorted' (P1). For students to 'imply that [staff] haven't seen something that [they] should have done, ... is really anxiety inducing' (P2).  

The potential for technology to support \textit{change at an institutional level} included more efficient triaging, providing information to support students in carrying out more of the advocacy themselves. S1 explains how they 'feel like I spend a lot of my time sending emails to various people and chasing things up and having to advocate for myself which is fine, but it's time consuming and I think if at least something like a robot could easily answer sort of basic questions at least like how a chatbot might, then that could be helpful. I think sometimes there's more nuance to it, where a human being would be better, but maybe it would be a way of sort of like triaging requests like some things would be easy to solve using that solve technology'.

However, there was also concern over attitudes to new technologies in a climate where students and staff are seeing the university pull back from technological change in other areas such as examinations and lecture delivery. P1 reflecting that 'all of the students are so used to learning technologies and are making use of them in really, really imaginative ways, and suddenly they're back in lectures where people are talking on boards and they're expected to hand write'.

\subsubsection{Authenticity in the Higher Education Context}
The participants stressed the importance of control over their own \textit{authentic voice}, particularly when using tools to clarify their communication. S1 explained that  'if I'm emailing someone for the first time or I don't know what sort of tone to strike. I think it's helpful to …put in my draft and then and then see what it says… sometimes it sounds completely inauthentic and I don't think that gets the sense of what I'm saying, but other times it's been really helpful'.

However, there was concern over the impact of the \textit{strangeness} of a robot in higher education contexts, preventing normalisation to the point of integration into student support.  P2 explained that 'For me, [as a neurodivergent person] I might be so distracted by ... its movement, the kind of visual aspect of it that very quick comparison that you're sort of doing between like it doesn't look human, but it's sort of trying to be', going on to say 'sound of a voice is really important for somebody to open up or kind of engage…those kinds of things combined with a neurodivergent brain would be so distracting'. Self-consciousness adds to the challenge of strangeness, with S1 describing 'I think I feel a bit self-conscious about interacting with it. I think especially in front of other people, it's hard to know … I'd feel like people would be thinking about how I was interacting with it, whether I was treating it like a human or or not. ..I think I'd find that tricky'. Participants made a distinction between forms of interaction: 'I don't like the fact that it's humanoid. I'd rather like talk to like an Alexa little box thing. I don't necessarily need it to look a bit like a human. I don't really see the point in that' (P2). With S2 explaining that 'I really don't like the way that they're made to ... simulate a human being ...  they're not good enough to actually be uncanny, but heading in that direction ... I would just find it really creepy and horrible'.

For S2 this was a philosophical issue; 'I guess my instinct is that I don't like humanoid robots because you can tell that it's trying to be like a human, but it's not. It's not a human, it's a machine. It has a sphere that's meant to be a head, and it has sort of limbs, but it's not [a human] And so it's not social interaction. At least I don't believe it is. It's not social interaction talking to ChatGPT. Talking to a humanoid robot is not talking to a person ...  To make a clear distinction between what is and isn't a person, it might sound silly, but we are in a time now where there is a lot of discussion of what is human intelligence, like, what is intelligence, what is not'.

Participants were optimistic about the normalisation of robots in higher education, drawing on recent experiences of the emergence of AI tools such as ChatGPT, and pandemic-related emergency remote learning.  Participant 1's view that 'it wouldn't be othering ... technology that has become normalised if everybody is using it, it's not seen as special' was echoed by Participant 4's view that '(robots technology) is a tool that is going to become the norm, and I think if we don't use it then (we) will be behind, so I think you just have to adapt to it'. This perspective was summed up by Participant 1 as ' unfamiliar rather than problematic'. 

Participants reported tensions and ambiguities in \textit{maintaining boundaries }between the individual, the institution and sources of support. The introduction of tools, such as robots, add to these existing issues, raising questions about advice or responses when the individual's needs are in conflict with the needs of the institution. The boundary between the student themself and the technology was expressed as concerned over legitimacy of the communication, with Participant 3 wondering 'in terms of …legitimacy … is this really coming from me?'. 

Where participants had not previously considered the potential for social robot use in supporting disabled students' advocacy, multiple issues were discussed around the \textit{complexity and nuance of human support} and how this could (or could not) translate into interaction with a social robot. 
Some participants expressed optimism over adaptation to robot technology: 'I'm sure it would have its quirks, much like being discovered with any form of new technology, you know the like, you can't pick up the sarcasm over text in the same way you might not pick up on some of the nuanced body language from the robot. But you learn to adapt depending on the medium that you are speaking to' (P1). Furthermore, social robot use represented a logical extension of existing technology in student support as 'other universities  have got chat bots set up in their student support unions that just ask those basic initial questions and arguably, even things like the student listening service Nightline.. were [based around predetermined] prompts ... I can't see any reason why that kind of prompting can't come from somewhere else' (P1). Social robots having an advantage over human support as 'that robot has oceans of time and is not judging' (P1).

When exploring the potential for social robot use with students, that use was contingent. For P2, social robots would be 'useful as long as there was a conversation that they had with a human ... prior to that or afterward', and for S2 if the information and advice coming from the social robot was 'genuinely quite accurate, then that would probably be useful'. Participants drew a distinction between communication that is 'something which ... is quite black and white or … very specific [to be communicated] in a more succinct way [and] something that's a bit more nuanced and [needs] to have more of a conversation around [it], so nuance [is a] big area' (P2). The human conversations draw on deep relationships that form between the student and support workers; 'you work together for a long period of time on a regular basis …  over that course of history they learn a lot about themselves and I’m just not sure whether that will be [possible for a robot]... those conversations, those nuances, those intuitive and often very spontaneous interactions that lead to huge revelations … that really deep, deep kind of learning that they can rely on that they can rely themselves' that take place with human support that would be beyond the scope of robot support' (P2). 

\subsection{Summary Phase 1}
In summary, through the interviews with disabled students, and experienced disabled disability professionals we found that issues around empathy are a key priority, with particular sensitivities in the higher education context. Disabled students commonly experience a lack of empathy in their interactions with peers, teaching and support staff in the university. Therefore, technological tools, such as social robots present the risk of reproducing and potentially amplifying biases in a more intimate setting, creating additional vulnerability for the disabled student. Furthermore, students and practitioners had concerns around privacy and the blurring of boundaries between the student and institution where experiences and concerns are disclosed and advice sought from a robot that is funded and provided by the HEI.

Whilst the \textit{strangeness} of social robots was considered a challenge, views diverged on whether this would be trivial to overcome through familiarity or whether social robots were inherently unsuited to support in this context. Given the dynamic and evolving nature of disability itself, participants reflected on the role of technology in this evolution of understanding. In this aspect, participants explored complex interconnections between the technological tools, institutional structures, and societal change. Specifically in the student cohort, new framings and language are embraced by students, such as that around neurodivergence, with participants speculating that increased use of technology could hasten and clarify language around disability. Despite this potential to support change, concerns remain around the authenticity of communications generated with the aid of a robot. These concerns were grounded in experience with existing AI tools used by students where clarity and directness of communication were seen to be at the risk of losing one's authentic voice. 

However, the potential for social robots to improve understanding was recognised across the participant group. Firstly, a role for social robots in providing a non-judgemental and predictable source of information and advice was welcomed. Secondly, a broader, institutional level role was imagined as part of wider institutional cultural change through greater sharing of information, clarity around disability access rights and shared, responsive language around disability.  In the higher education context in particular, language preferences and expectations evolve rapidly as new cohorts of students enter the university community, so this updating role was seen as potentially useful dimension for social robot in the university context. Power dynamics emerged as important, with students seeking to maintain agency, control and autonomy both within the current structures and in terms of the potential future use of social robots.

After concluding Phase 1, the results from the interviews were synthesised to inform the second phase of the research project: Phase 2 conducting focus groups with groups of disabled students to further ideate and envision how social robots can be implemented for mediation and advocacy for students with disabilities. The next section account for the methodology and results of Phase 2, followed by a discussion of both Phase 1 and 2.

\section{Phase 2: Focus Groups}
The focus group phase consisted of two separate focus groups in which the content and presentation items were, to a large extent, informed by the interviews in phase 1. More specifically, the themes arising during the interviews, such as examples of existing technology usage, perceptions of technological and AI authenticity, empathy, and normativity were brought up also in the focus groups. These themes were distilled by initiating the thematic analysis of the interviews while planning the topic guides for the focus groups. This section summarises the methodology of the focus groups, and the results are presented in the subsequent section of this paper. A glossary of terms was provided for all participants in both focus groups and can be seen in Appendix A.1. 

\subsubsection{Participants}
Five participants were initially recruited for the first focus group session. Of these, four participants also participated in the subsequent focus group session, ensuring continuity in the discussion while allowing for the exploration of evolving perspectives. All participants were students at the University of Cambridge with a self-declared disability (either recognised by the University or not) recruited from colleges and departments via their mailing lists. All students were studying at a second-year undergraduate level or higher, to ensure they had experienced accessing the university's disability services over time. All students were from the School of Humanities and Social Sciences, for the reason that lab adjustments can be quite specific, so broadly similar essay-based subjects were chosen so that the participants could have some commonality of experience to bring to the discussion. The first focus group had 1 male, 1 non-binary, and 3 female participants, between the ages of 20-43 (median 21), diagnosed with either ADHD or Autism as self-declared and recognised by the University of Cambridge. The second focus group had 1 male, and 3 female participants. The first focus group was 1 hour and participants were compensated with an Etsy or Amazon voucher of £12, the second focus group was 1.5 hours (based on the results of the first focus group requiring more time), and participants were compensated with a £18 voucher. 

\subsubsection{Setting} 
The focus groups were carried out in person (except one participant joining remote in the first focus group). Those who participated in person met in a seminar room at the University of Cambridge, and a Microsoft Teams meeting was set up to enable discussion as well as voice recording and transcription. The focus groups used the online interactive presentation software Mentimeter \footnote{https://www.mentimeter.com/}, to enable participants to interact with the prompts, write down ideas and thoughts in real time anonymously through their own phones or laptops, put virtual stickers on virtual post-its and rate how they agreed or disagreed with the statements. They could therefore see each other's responses anonymously in real time. The results derived from the Mentimeter interactive presentation can be seen in the results section. The focus groups were both video and audio recorded.

\subsubsection{Procedure}
The agenda of the first focus group revolved around three main focus areas. The first was related to \textit{understanding the needs and problem space of students with disabilities}, including the existing support for mediation and advocacy, usage of technological tools and the existing barriers, and successful features. The second part related to \textit{ideating robotic support for disabled students} and introduced the concept of social robotics as a potential supportive technology, and aimed at understanding the participants initial perceptions of the technology. This part was initiated with showing the participants a short video clip of social robots\footnote{https://www.youtube.com/watch?v=cF8OuaY81vg} without using the sound, but just to visually present the most common social robots to create a unified understanding of what they look like and how they operate. The third focus area revolved around the potential features, limitations, and capabilities of using social robots for mediation and advocacy for students with disabilities. A breakdown of the structure of the first focus group can be seen in Table \ref{st}. 

\begin{figure}
    \centering
    \includegraphics[width=1\linewidth]{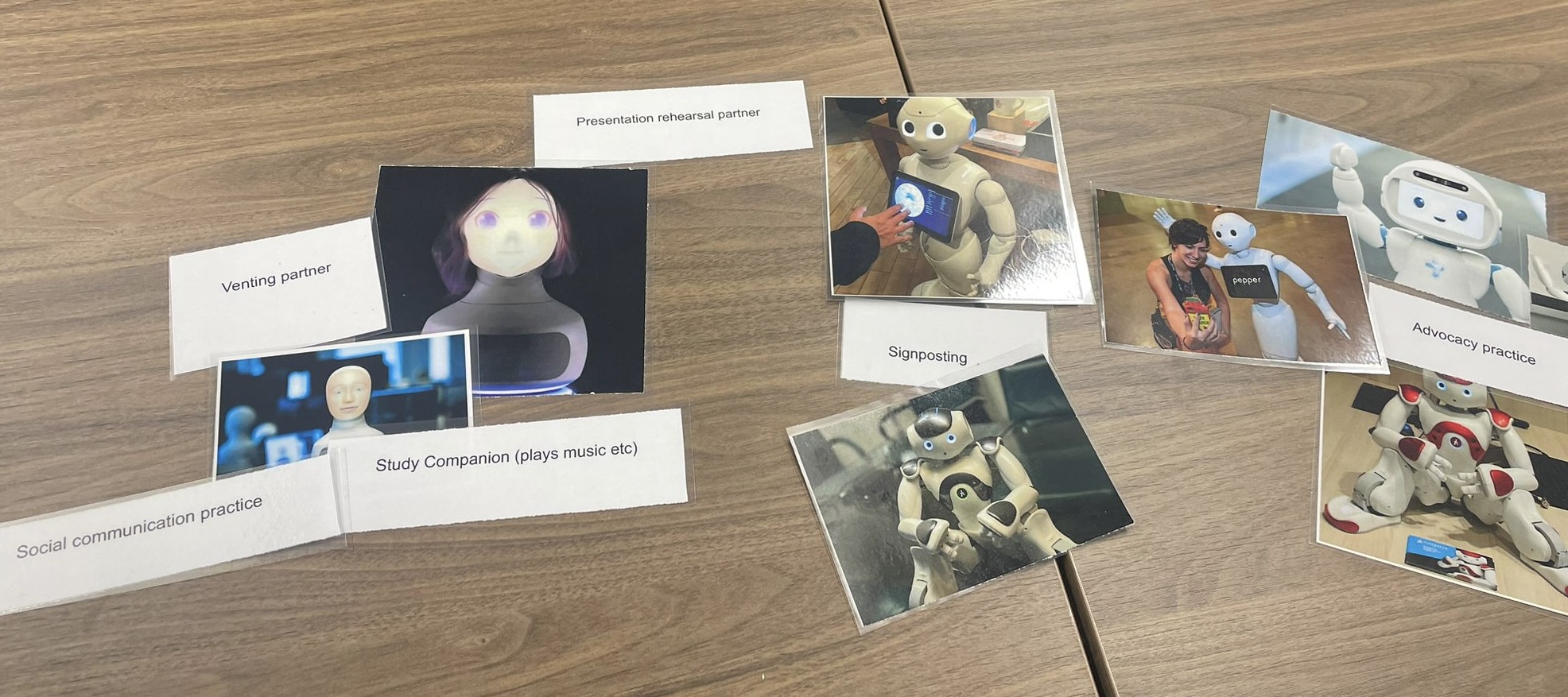}
    \caption{Interactive activity in the second focus group in phase 2 of the study. Participant's matched examples of robot roles derived from the first focus group with pictures of different robots}
    \label{fig4}
\end{figure}

The second focus group was related to more specific design considerations and roles of social robots. It also included a short demo of a QT robot (LuxAI)\footnote{https://luxai.com/} interacting with one of the researchers at the start of the focus group to allow participants to see its potential capabilities. The demo included the first part of the VITA system, a social robotic wellbeing coach introducing itself, as described in \cite{spitale2023vita}. The remainder of the focus group was related to different dimensions of the robot design. The first focus area concerned the robot role, and participants got a selection of print-outs with pictures of different social robots (Pepper, Nao, Furhat, and QT) that they matched with print-outs of potential roles, such as 'Venting Partner' or 'Signposting' (see Figure~\ref{fig4}). The roles were derived from the interviews and the first focus group as examples of potential use cases of robots students had mentioned during the first focus group. One of the focus areas from focus group 1 was to envision the social robot's potential role, what the students want from the robot, where it should be located, what they need it for and what topics they would want to talk about with it. Commonly occurring replies to these questions were distilled into specific roles used in the second focus group. The second focus area was related to the design and envisaging of physical and non-physical characteristics of the robot itself, such as its form, voice, appearance, and identity. The last focus area was related to the design and envisaging of contextual, situational and structural factors of the robot, such as the participant's relationship to it, robot's environment, terms of service and consent. These contextual design dimensions have previously been investigated for algorithmic systems \cite{rakova2023terms} as well as for social robotics specifically \cite{markelius_design_justice, axelsson2021social, ostrowski2022ethics} to follow principles of Design Justice \cite{costanza2020design}. A detailed overview of topic guides of the focus groups can be found in Appendix A.3-4. 

\begin{table}[h]
    \centering
    \begin{tabular}{|l|c|}
        \hline
        \textbf{Focus Group 1 Focus Areas} & \textbf{Duration} \\ \hline
        Warm-up and housekeeping & 10 min \\ \hline
        Understanding needs and problem space & 10 min \\ \hline
        Ideating robotic support for disabled students & 10 min \\ \hline
        Features, limitations and capabilities & 20 min \\ \hline
        Wrap up and reflections & 10 min \\ \hline
        \textbf{Focus Group 2 Focus Areas} & \textbf{Duration} \\ \hline
        Introduction and housekeeping & 10 min \\ \hline
        Social robot demo & 10 min \\ \hline
        Exercise with robot role print-outs & 30 min \\ \hline
        Robot characteristics design dimensions & 20 min \\ \hline
        Robot contextual design dimensions & 20 min \\ \hline
    \end{tabular}
    \caption{Structure and durations of the two focus groups}
    \label{st}
\end{table}

\subsubsection{Data Analysis}
A thematic analysis for each focus group was conducted following the 6-step thematic analysis methodology \cite{braun2006using} as well as the previously mentioned guidelines for rethinking barriers and enablers in qualitative health research \cite{haynes2024rethinking} to enable identification of structural and nuanced issues. The data used in the method were the transcribed audio data from both focus groups, as well as the text inputs in the Mentimeter. The quantitative inputs in the Mentimeter were analysed separately. This approach has previously been used for iterative design and development of social robotics \cite{axelsson2024robots}. The six steps are as follows: i) \textit{Familiarizing with the data} this involves transcribing the data, reading through it, and writing down initial observations; ii) \textit{Generating initial codes} identifying codes within the dataset and organizing data under the corresponding codes; iii) \textit{Searching for themes} grouping codes into themes and gathering all relevant data under each theme; iv) \textit{Reviewing themes} ensuring that the identified themes align well with the associated codes; v) \textit{Defining and naming themes} developing clear definitions and labels for each theme, ensuring coherence with the overall narrative of the dataset; and vi) \textit{Producing the report} selecting examples to illustrate each theme effectively. The analysis was a hybrid approach, a combination of deductive and inductive coding in line with \cite{fereday2006demonstrating} to allow for a rigorous and comprehensive identification of themes. The topic guide (Appendix A.3) served as a scaffolding for the deductive coding.

\subsection{Results}
This section accounts for the qualitative results of the study, as well as the quantitative data collected through the Mentimeter during the focus groups. The thematic analysis was carried out in three parts, one covering the four interviews, and one for the two focus groups each, which are here accounted for in one section each respectively.  

\begin{figure}
    \centering
    \includegraphics[width=1\linewidth]{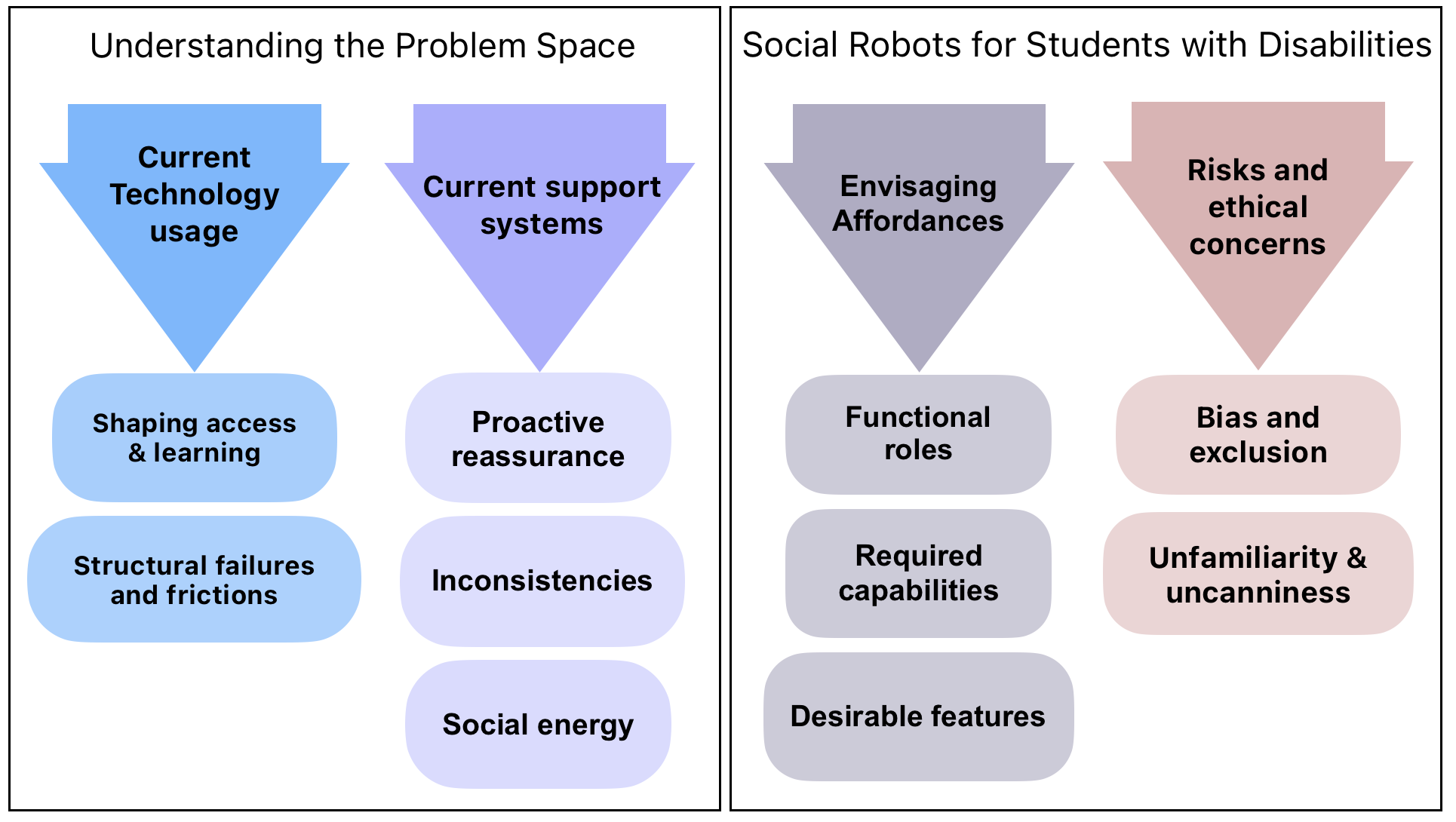}
    \caption{Overview of the themes and subthemes derived from focus group 1}
    \label{theme2}
\end{figure}

\subsection{Focus Group 1}
Both qualitative and quantitative data were collected from the first focus group. Quantitative data was collected through the Mentimeter, and are presented in section 4.2.1. The thematic analysis of the qualitative audio recordings from the first focus group resulted in two overarching parts: \textit{understanding the problem space}, and \textit{social robots for students with disabilities}, each with their respective themes and subthemes. An overview of the themes can be seen in Figure \ref{theme2}. The qualitative data are presented in section 4.2.2-3.

\begin{figure}
    \centering
    \includegraphics[width=0.75\linewidth]{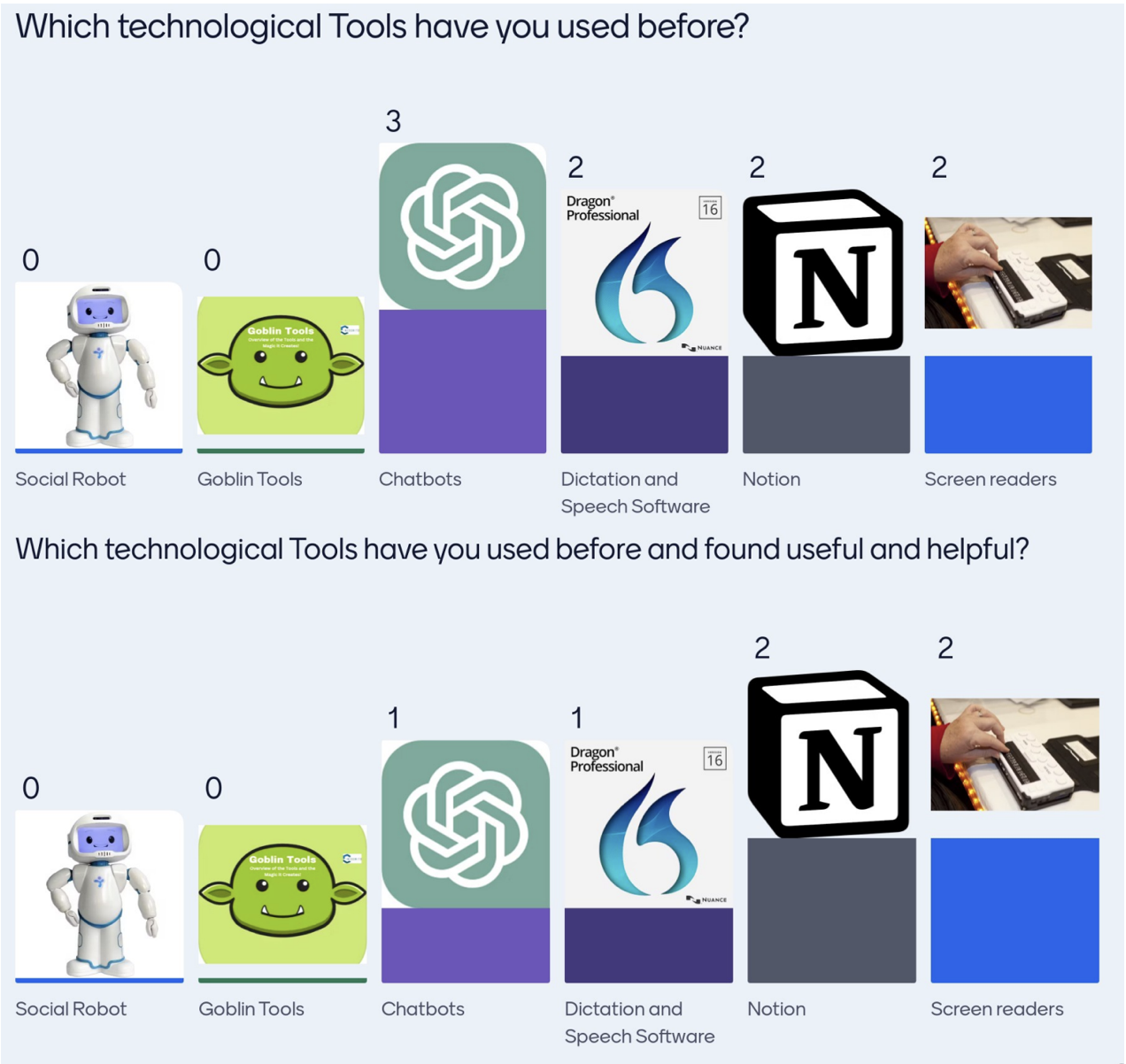}
    \caption{First task in the focus group was to indicate existing usage of technological tools}
    \label{tools}
\end{figure}

\begin{figure}
    \centering
    \includegraphics[width=0.5\linewidth]{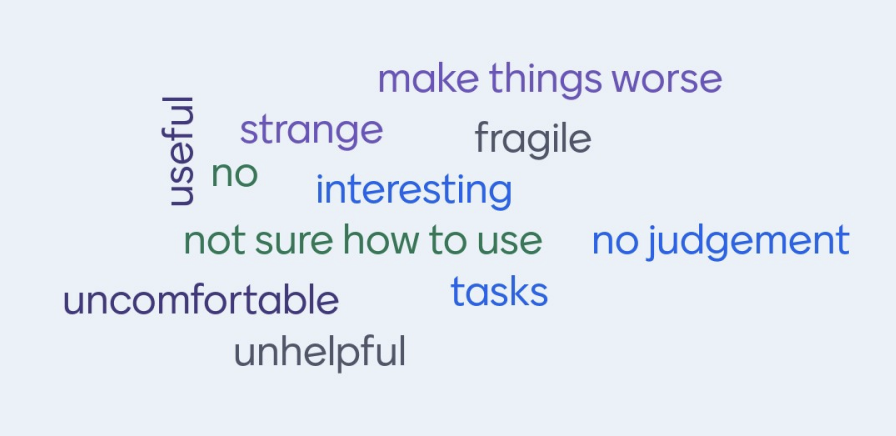}
    \caption{A word cloud from the question 'What are the first words coming to mind after seeing these robots?' after showing the video clip of social robots during the Ideating Robotic Support part of the focus groups}
    \label{words}
\end{figure}

\subsubsection{Mentimeter Quantitative Results}
\begin{figure}
    \centering
    \includegraphics[width=1\linewidth]{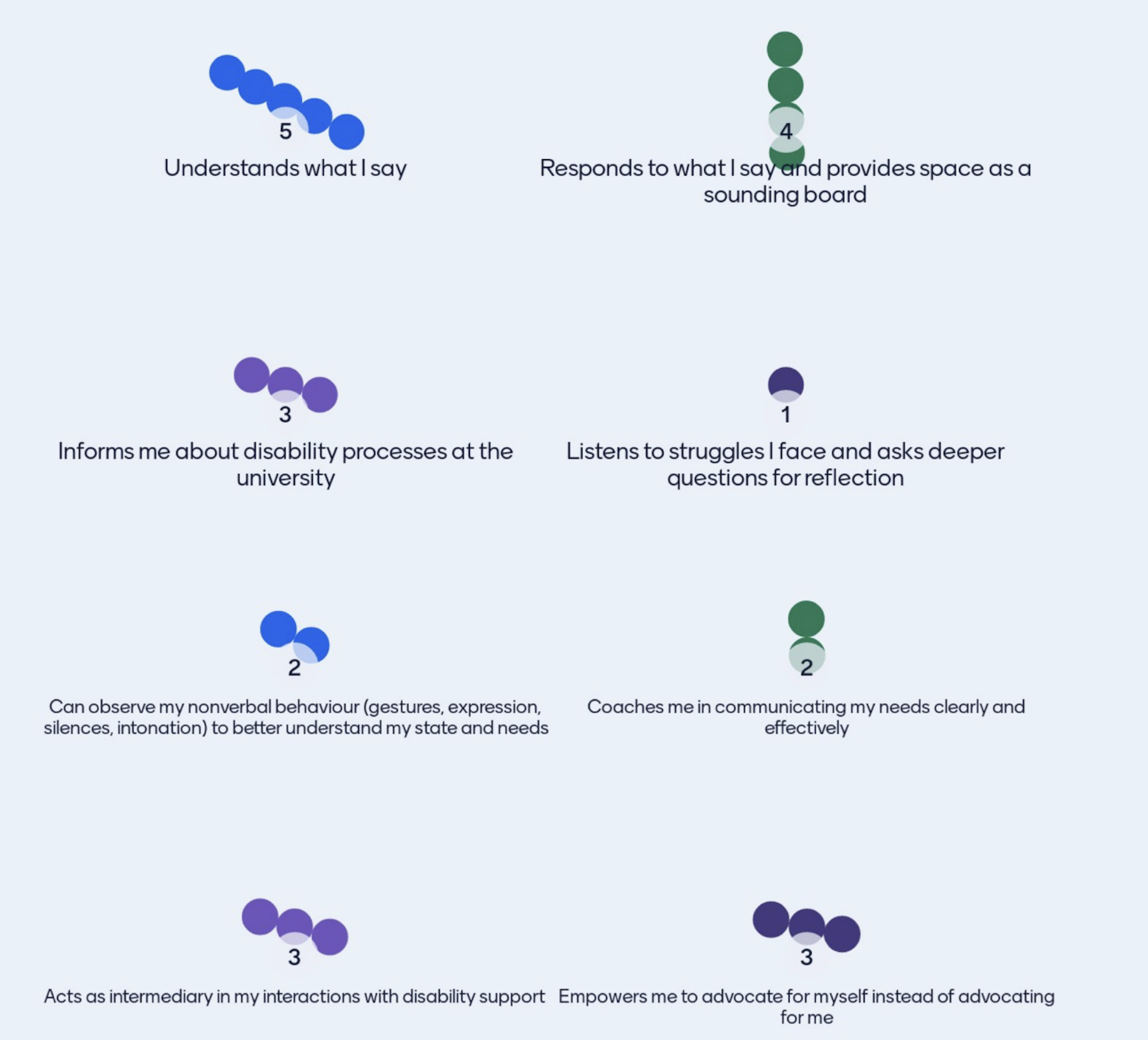}
    \caption{Virtual post-it responses to the question 'Which robot capabilities are most important to you?' in the first focus group}
    \label{fig}
\end{figure}

\begin{figure}
    \centering
    \includegraphics[width=1\linewidth]{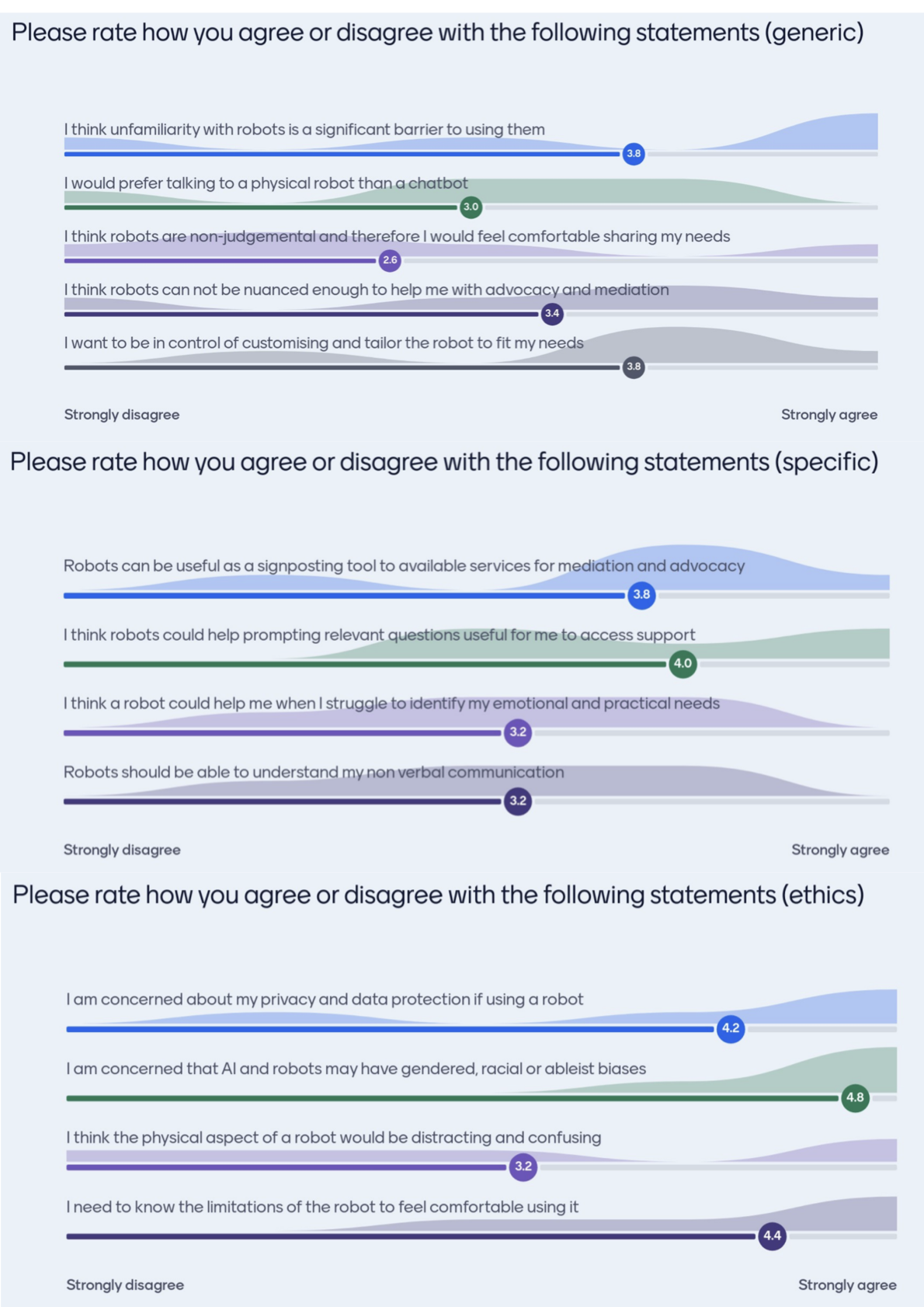}
    \caption{Ratings related to focus area 3, specific features, limitations and capabilities of social robots in the first focus group}
    \label{rate}
\end{figure}

\begin{figure}
    \centering
    \includegraphics[width=0.75\linewidth]{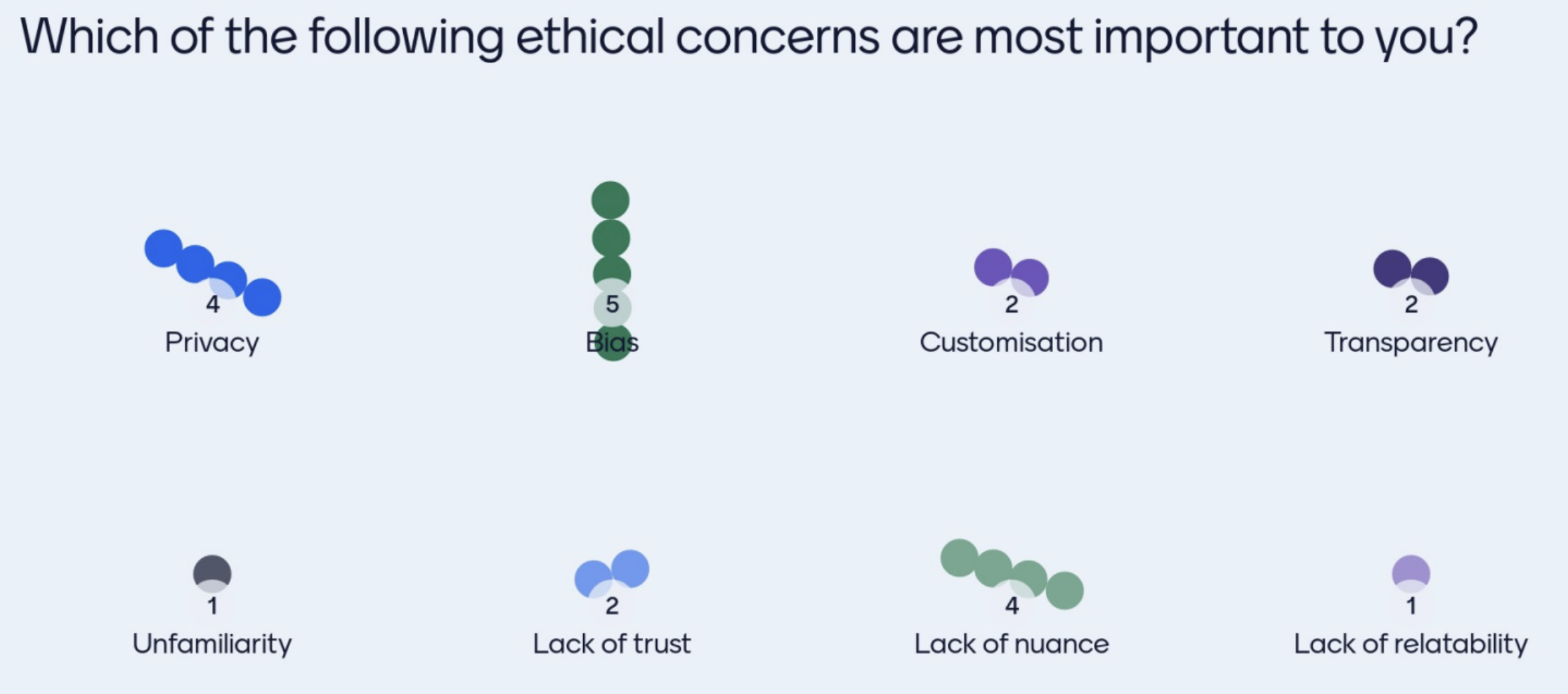}
    \caption{Virtual post-its related to focus area 5, ethical considerations and limitations, in the first focus group}
    \label{eth}
\end{figure}

In the first focus group, the Mentimeter presentation software was used to collect quantitative findings. The first part of the focus group was a warm up exercise related to the general perceptions of technology, AI and robotics where the findings can be seen in Figure \ref{tools}. In focus area 2, related to ideating robotic support for disabled students a word cloud was created with the students' initial reactions to seeing a video of social robots. This word cloud can be seen in Figure \ref{words}. In focus area 3, specific features, limitations and capabilities of social robots were discussed together with virtual 'post-its' and ratings were used. In Figure \ref{fig} the virtual post-its are shown specifically related to robot capabilities and with regards to which capabilities are rated most important. This focus area also included ratings submitted through the Mentimeter, as seen in Figure \ref{rate}. Finally, in focus area 5 with the focus on ethical considerations and limitations, there was a final virtual post-it for participants to indicate which ethical considerations were most important to them as seen in Figure \ref{eth}. 

\subsubsection{Current Technology Usage}
The participants were then invited to discuss the current use of technological tools, including how they perceive it to be \textbf{shaping access and learning}. Most students had at some point during their university time used screen readers, or various forms of dictation and support for processing text. Students described the benefits of screen readers, initially using them as a way to multitask but discovering that combining audio and visual stimuli helped them focus more and identify key points in texts. They noted that listening while reading prevented them from accidentally skimming over important details, as hearing the words reinforced their significance. Customisation, such as adjusting tone, pitch, and voice type, was particularly valuable. For example, one student found deeper voices more engaging. Another student expressed fascination with chatbots, calling them “miraculous” for providing interactive ways to explore ideas. While acknowledging that chatbots like ChatGPT were often incorrect, they viewed these errors positively, as they prompted deeper learning and critical thinking. This student emphasised that conversational interaction, rather than solitary reading or writing, was their most effective way of learning, and chatbots offered a dynamic way to engage with new ideas.

Although some students found that these technologies increase both engagement, focus, and accessibility to important learning material, a range of \textbf{structural failures and friction} were mentioned. One student shared their frustration with screen readers, noting that while they initially believed the tool would suit their preference for auditory learning, the unnatural cadence and tone of the voices disrupted their focus and comprehension. They found the robotic reading style confusing, making it difficult to follow sentences as they would with a human voice. Another student expressed a similar struggle, explaining how the voices in screen readers clashed with their internal reading voice, making it impossible to engage effectively with the text. Students also mentioned friction in their experience using chatbots and screen readers for language learning. They wished they had known about suitable screen readers in other languages than English, highlighting a gap in available tools for multilingual learners. Similarly, they found interacting with chatbots challenging due to a lack of clarity about academic policies and discomfort with prolonged "conversations" with AI. Please see examples of specific quotes in Table \ref{qou}.

\begin{longtable}{|p{0.999\linewidth}|}
    \hline
    \textbf{Please provide examples of when mediation and advocacy was successful} \\ \hline
    "[Accessibility and Disability Resource Centre] was invaluable in their support to create the SSD which was provided to all academics." \\ \hline
    "The SSD making it possible to submit other forms or work (plans instead of essays) throughout the year was really helpful."\\ \hline
    "I reached a point when I felt exhausted, and wasn't sure if I should go to my lessons anyway, but didn't have enough energy to check with the teacher. My [director of studies] mediated by emailing the teacher to explain." \\ \hline
    "Emailing the college to get allocated a room with a bathroom for disability reasons was very successful." \\ \hline
    "My mentors were invaluable in helping me stand up for myself too." \\ \hline
    "The creation of disability families in college was helpful because it made me feel understood and seen." \\ \hline
    "The [Disabled Students' Allowance] assessment was helpful to verbalise and understand needs especially when newly diagnosed." \\ \hline
    "I could get exam accommodations here which I had needed at school but hadn’t received - the uni seemed more open to the idea of exam adjustments than my school with better systems in place." \\ \hline
    "Requesting an SSD was somewhat successful. They initially misplaced my information, but being able to record the lectures and get exam arrangements has been invaluable." \\ \hline
    "Directions/clarity on exactly how to get admin stuff done (who to contact, how long things take)." \\ \hline
    
    \textbf{Please provide examples of what is missing for mediation and advocacy efforts} \\ \hline
    "An accessibility first approach. It is still seen as an add-on or special arrangement." \\ \hline
    "Despite having on my SSD that all lectures should be recorded, some of my lecturers still refuse to record them, having someone push for recordings." \\ \hline
    "Consistency - departments can respond quite differently to SSDs or requests for adjustments and this can be stressful as some students need to self-advocate more than others." \\ \hline
    "The ADRC and the college both provide help for disabled students, but it's really hard to tell who to contact if you need it. At least the SSD reduces the need to go back and forth between these." \\ \hline
    "It would help to streamline the process if there was less ambiguity about which people to email and which forms to fill out to get accommodations." \\ \hline
    "Some help identifying and communicating my needs in a range of contexts. I find this particularly hard in relation to students who lack information about my condition, e.g., people in shared kitchens."\\ \hline
    "Some things like access room ballots are much later or less clear than their non-access counterparts - this can put students who need them at a disadvantage or make them harder to access." \\ \hline
    "Awareness training for all gatekeepers such as facilities and porters." \\ \hline
    "More buy-in from the facilities department to allocate resources such as quiet spaces." \\ \hline
    "In some cases, user-friendly systems for getting adjustments. The ADRC is great but can be difficult to find out how to use it without help from older students." \\ \hline
    "Performance standards for college senior tutors. Having it as part of their job description and performance review to support accessibility." \\ \hline
    \caption{Quotes from the Mentimeter in focus group 1 related to understanding the problem space for students with disabilities in higher education}
    \label{und}
\end{longtable}

\subsubsection{Current Support Systems}
Having discussed current usage of technological tools, the second theme related to understanding the problem space touched on current mediation and advocacy support practices at university. The Mentimeter entries can be seen in Table \ref{und}. \textbf{Proactive reassurance} emerged as one of the most important themes relating to structural and relational features of the current support systems. Students emphasised the importance of proactive reassurance provided by the accessibility services at their university. They described how the Accessibility Resource Centre (ADRC) created a comprehensive Student Support Document (SSD) outlining their diagnosis and the accommodations needed, such as longer deadlines, recorded lectures, and meeting agendas in advance. This document was automatically shared with lecturers, ensuring that accommodations were implemented without requiring the student to repeatedly disclose their needs or justify their requests. Students explained how this alleviated feelings of doubt and judgment often associated with self-advocacy. They appreciated how the system reframed accommodations—not as special favours, but as essential measures to ensure equal access and enable them to fully participate in their academic experience.

At the same time, students expressed a range of \textbf{inconsistencies} related to accessing disability support services, such as misplacing of documents, or having to chase and initiate contact with the university to get support. For example, one student explained inconsistencies in the support provided by disability services and departments, particularly in relation to accommodations. They observed that while some peers received the support they needed, their own department presented significant challenges. For instance, the department did not record lectures by default, despite student requests for this accommodation. Students were often told to handle recordings themselves, but accessing the necessary equipment or guidance on its use proved difficult. The student described how even attempts to seek help from central services, such as the ADRC, led to confusion and unresolved issues. These inconsistencies show that the availability and implementation of accommodations varies greatly across departments, leaving students uncertain about whether essential support would materialise in practice.

Finally, the last subtheme concerns \textbf{social energy} in relation to experiences of current support systems. Although contact with relevant university staff, such as advisors, directors of studies and professors were perceived as mostly successful, students also found them to be socially tiring, and demanding a significant amount of social energy. Inconsistencies were also mentioned here, as different university personnel often were found to be varying in the support they offered. Also related to the social energy, one student described the challenges of navigating non-academic spaces and interactions as a disabled person, and the absence of structured support in those contexts. They noted that shared spaces, such as kitchens or casual conversations with peers, often lacked awareness and sensitivity regarding disabilities. They also reflected on the ongoing process of learning to self-advocate in such situations, expressing how the absence of someone to mediate or provide support added to the difficulty. While these challenges did not directly affect their academic performance, they emphasised the need for support mechanisms addressing issues that, while not strictly academic, still had a significant impact on their overall experience and time management. Please see examples of specific quotes in Table \ref{qou}.

\begin{longtable}{|p{0.1\linewidth}|p{0.8\linewidth}|}
    \hline
    \rowcolor[gray]{0.9} 
    \multicolumn{2}{|c|}{\textbf{Understanding the Problem Space}} \\ \hline
    \textbf{Current Support Systems} & 
    "\textit{[SSD] a big document which says 'this is her diagnosis and this is all the things that would help her.' Longer deadlines, recorded lectures. Agendas for every meeting beforehand, all that kind of stuff [...] So that's really helpful because a lot of my feeling about it was that it always feels like you are doubted and that you have to justify yourself.}" \newline
    "\textit{Things like the SSD are very inconsistent with different people [...] If you go ask for something, does that mean you're actually going to get it? There are limits in the consistency for those accommodations actually mean in practise.}" \\ \hline
    \textbf{Current Technology Usage} & 
    "\textit{I would have really appreciated if there had been a screen reader available in Arabic that I'd known about. With ChatGPT, it was recommended for translation homework and I wasn't sure I found it helpful. Because I ended up having a conversation with the chatbot saying I'm not sure if I trust this. [...]  I didn't really feel comfortable having a conversation, which wasn't a conversation with a robot for about two hours.}" \newline
    "\textit{I find chat bots just miraculous and so interesting, mainly because they're generally wrong [...] I'm a learner through talking about things and exploring ideas, that's where I get my energy from. So writing and reading alone has never been a good way of me learning so it is just brilliant to have an interaction. But I know they're not always right, but that actually feeds into my desire to learn and to to find out the right answer.}" \\ \hline
    \rowcolor[gray]{0.9} 
    \multicolumn{2}{|c|}{\textbf{Social Robots for Students with Disabilities}} \\ \hline
    \textbf{Envisaging Affordances} & 
    "\textit{Sometimes I do need to talk something out. I'd just rather not do it with a person because that takes a lot of effort to maintain social cues and to not say too much. And so if I needed to talk about something, or figure something out for myself, it might be nice to talk to something that was replying back to me, but I knew it wasn't perceiving me. It was just replying. That might be OK because it might use less energy.}" \newline
    "\textit{A really valuable feature in my opinion would be the chance to give continual feedback to the robot and/or its developers – contributing to the creation of more effective personalised support. I believe that understandings of disability are constantly under development (in terms of medical definitions, and the ways that individuals conceptualise their conditions – for example, the “social model” of disability has made a huge difference in helping me to overcome stigma around autism), and robotic support would ideally reflect this exciting state of flux.}" \newline
    "\textit{I think to to feel like I actually want to share my needs, I probably need to have a sense of what I was gonna get back from it [robot] because it would either be to make myself more aware of what I need or to help me find a way to express that.}" \newline
    "\textit{A final consideration is that I shouldn’t let anyone else be the expert on me! I think it is important to acknowledge that nobody, human or AI, has all the solutions to the difficulties we face. It seems to me that positive experiences of seeking solutions are really the end goal here. Could the inevitable non-expert-ness of supportive AI be expressed in such a way that students feel empowered to foreground our human experiences?}"
    \\ \hline
    \textbf{Risks \& Ethical Concerns} & 
    "\textit{I think some AI software right now are quite judgmental. I also think if you're going to a robot for advice in your life, the way that they're going to be able to advise you is by looking at stuff online, a lot of the stuff is written by the same all dead white men. So for me as a person of colour and someone who is trans and queer, it would be not very applicable.}" \newline
    "\textit{My first reaction to robots is probably quite a negative one, because it always seems to take me a very long time to understand what something is offering me so that I can evaluate if I want to use it or not. [...]. But also I've had some quite difficult experiences of advocating for myself, I also have had a lot of nice, very human conversations which you felt quite transformative for the way I think about myself and my disability, and I'm not sure that a robot could offer that nice feeling.}" \newline
    "\textit{A lot of the things that I read about autism are only just beginning to include people from more than one demographic. And so sometimes you have to do extra special care to check that what you're reading has been including people of colour, femme presenting people and queer people. And because that often isn't the case, and because you have to actually look to find whether or not the research included that, or even considered it, I don't think a robot would do that. Unless it was strictly told. And even then it might not always do it right. And that could really impact the bias of the robot.}" \newline
    "\textit{The interface of the robot might not even be made to interact with autistic people, if the robot was built to say, only talk to people who were looking at it in its face, then that might be problematic. I can think of a bunch of other things for other neurodiverse groups where the robot would have to be made with them in mind in order for it to actually work, because there's a lot of assumptions about what communicating is, and if the robot was built on those, not everyone would be able to even use it.}" \\ \hline
    \caption{Quotes from research participants in the four main themes identified in the study}
    \label{qou}
\end{longtable}

\subsubsection{Envisaging Affordances}
In this part of the focus group, students were able to freely envisage how they could see social robots taking place in their lives and supporting with mediation and advocacy. Firstly, students discussed different \textbf{functional roles} of the robot and expressed a need for robots to fulfil specific, manageable tasks that alleviate the burden of social interactions or complex communication. For instance, one student envisioned a robot that could provide responses without the emotional or social labour typically required in human interaction, and would serve as a low-energy solution for processing thoughts or figuring out problems independently, without the need for social cues or maintaining conversations. Moreover, the robot was envisioned to assist with smaller, well-defined tasks rather than replacing human roles. There was an emphasis on the robot's ability to support specific aspects of tasks, like time management or communication, without overstepping into areas where human engagement is necessary. The functionality of a social robot was also framed around its ability to save time and energy, particularly in areas of communication and self-advocacy. Lastly, the role of the robot in creativity or personal expression was also considered. While not fully realised, some students expressed interest in robots that could enable creative engagement, like assisting in art or providing personalised feedback in interactions.

Students went on to discuss \textbf{required capabilities} of social robots. Many expressed a need for robots to understand and interpret the user's needs with a high degree of accuracy, such as recognise when they needed assistance, even if they themselves were unaware or hesitant to ask for help. This robot would ideally be able to take action on their behalf, such as reaching out to someone who could provide support. Students highlighted the importance of a robot’s ability to personalise interactions and integrate into everyday tasks, such as playing music that suited their mood or needs, and offering small gestures of support like telling a joke or playing a favourite song. The physical presence of a robot also came into play when imagining its capabilities. While some expressed concerns about the discomfort of having a physical robot in their living space, others saw value in having a robot move around their environment, like a “college pet”. Students saw the ability to provide feedback to the robot and its developers as a critical capability. This would allow disabled students to shape the robot's performance and effectiveness in meeting their needs. The opportunity to contribute to the ongoing development of personalised support was seen as valuable, particularly in light of the evolving understandings of disability. Students emphasised that robotic support should reflect the dynamic nature of disability, acknowledging how definitions and perceptions are constantly shifting. This feedback loop would ensure that robots remain adaptable and responsive to the changing needs of disabled people. 

Finally, in envisaging affordances, students went on to discuss \textbf{desirable features} in the robot. One of the most prominent features mentioned was the robot’s appearance. Cute designs were seen as a crucial factor in order to be approachable and effective in its role, with participants suggesting that without a friendly or appealing aesthetic, the robot would fail to engage its user effectively. Beyond cuteness, many participants expressed a preference for robots that did not try to mimic humans too closely. The idea of robots being human-like often felt unsettling, and many respondents pointed out that robots should embrace its own identity rather than trying to pass as people. Some suggested that robots could adopt shapes and features that were entirely new, potentially drawing inspiration from familiar objects or creatures, such as pets. The idea of a robot resembling a dog or a cat, each offering specific emotional benefits like companionship and comfort, was a widely appreciated concept. The design of familiar yet non-human objects, like the spherical and beeping BB8 or the "mouse robot" from Star Wars, was highlighted as another possible approach as they were perceived to have a desirable simplicity and ability to be non-threatening, yet endearing. The presence of sound or motion, even without a face, could make a robot feel alive and engaging. The robot's capacity to generate a positive emotional response—whether through cuteness, soothing sounds, or playful actions—was seen as essential for building trust and comfort in its use.

\subsubsection{Risks and Ethical Concerns}
The second theme related to social robots for students with disabilities is risks and ethical concerns. One of the key concerns raised by participants was the potential for \textbf{bias and exclusion}. A prevalent issue identified was the risk that social robots, like current AI systems, may perpetuate biases that reflect narrow, homogenous perspectives representing a limited demographic, predominantly cisgender, white, and male. For participants from marginalised groups, such as people of colour, trans, or queer persons, this was seen as a major concern that would make the advice or support provided by robots irrelevant, if not outright harmful. The issue of inclusivity was also raised with regard to the way robots might be designed to interact with different communities, particularly related to neurodiversity. Several participants noted that current research on autism and disability tends to exclude the needs and experiences of various marginalised groups, including people of colour and those with less visible disabilities. For example, the design and interface of the robot itself could exclude certain users, e.g. through the use of eye-tracking technology that expect direct eye contact could be problematic for neurodiverse users who do not find such interaction intuitive or comfortable. Failure to accommodate these communication aspects could make the robot ineffective or even harmful through reinforcing the notion that communication and interaction should conform to a narrow set of standards. These concerns were further emphasised by comparisons to biases found in other AI systems, such as image generation tools displaying racial and gender biases. 

Finally, ethical concerns and risks were also discussed in relation to \textbf{unfamiliarity and uncanniness}, particularly when robots attempted to resemble humans but failed to fully achieve this likeness. The "uncanny valley," was mentioned as a source of discomfort for many participants. The idea of robots having faces without being human made them feel strange or unsettling, leading to a lack of trust. Several participants felt that robots attempting to replicate human characteristics often failed, making them appear strange rather than helpful. Students mentioned it might also lead to awkward situations where others would question the use of humanoid robots which could reinforce stigma and making the user feel vulnerable or self-conscious about their need for assistance. The inability to fully understand the purpose or function of the robot added to this hesitation, as students felt they needed to spend significant time figuring out how to incorporate the robot into their lives and routines. A common theme was the difficulty of self-advocacy, which was often compounded by previous negative experiences in trying to get help. Participants highlighted the importance of human interaction in their lives, particularly in discussing their needs and disabilities. Transformative, human conversations had previously helped them better understand their own experiences, something they felt robots could not yet replicate. While some participants acknowledged that future robots could potentially improve and become more trustworthy, they were still unsure whether current technology could offer the same level of empathetic support or understanding.

\subsection{Focus Group 2}
\begin{figure}
    \centering
    \includegraphics[width=1\linewidth]{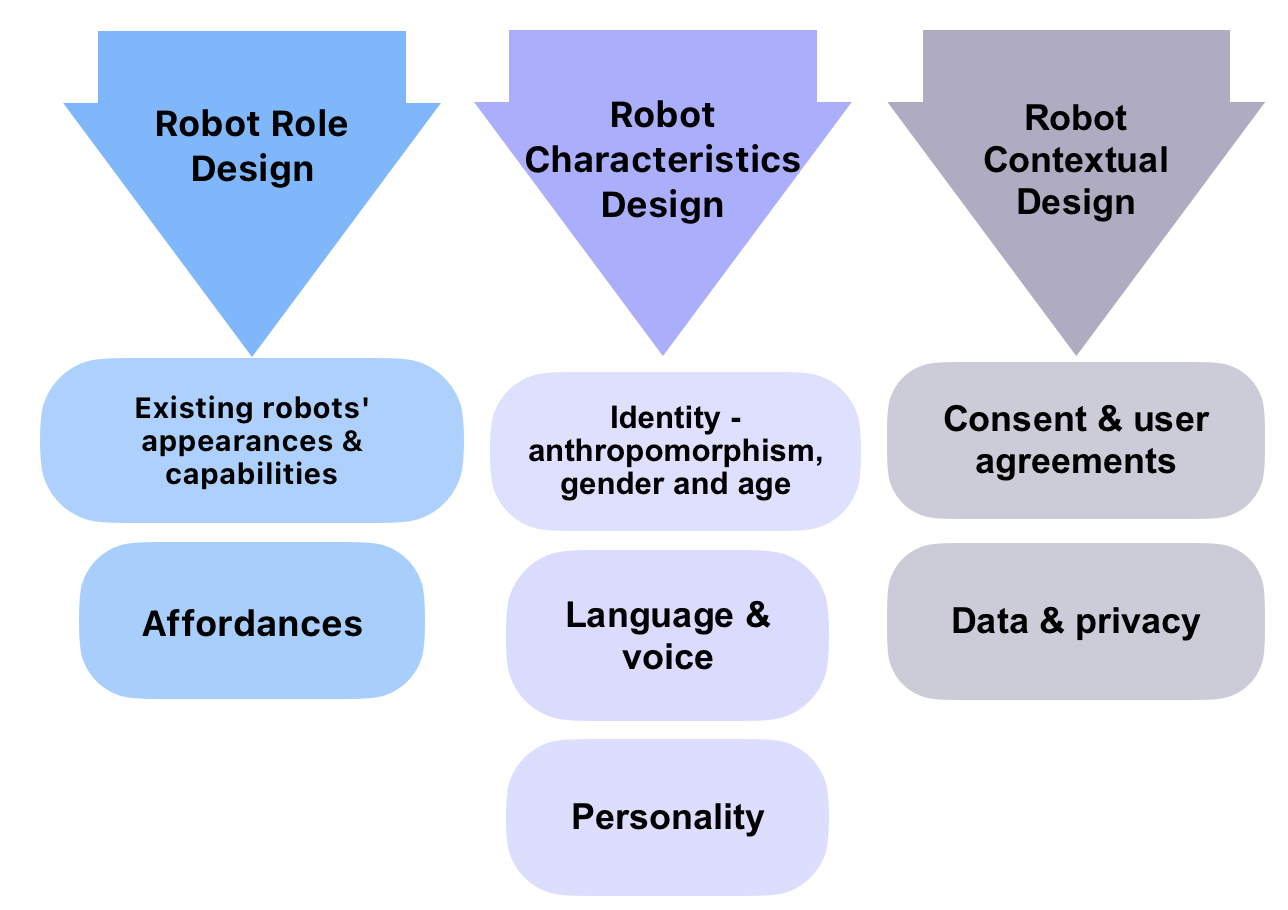}
    \caption{Overview of the themes and subthemes derived from focus group 2}
    \label{fg2}
\end{figure}

The second focus group involved four of the five participants that took part in the first focus group, a couple of weeks later. The main focus of the second focus group was on co-design, and to have a session with a stronger focus on designing and envisaging more specific aspects of a social robot to help mediation and advocacy for disabled students. After the exercise where participants got to match roles with robots, as seen in Figure \ref{fig4} the more reflective co-design started. An overview of the themes and subthemes from focus group 2 can be seen in Figure \ref{fg2}

\subsubsection{Robot Role Design}
The first discussion was centred around designing the role of the robot, based on ideas and envisaging made in focus group 1. These ideas were summarised into six potential roles: i) Venting partner, ii) Social Communication Practice, iii) Study Companion (plays music etc), iv) Signposting, v) Presentation Rehearsal Partner and vi) Advocacy Practice. The participants discussed these 6 roles in relation to different existing social robots, specifically Pepper\footnote{https://corporate-internal-prod.aldebaran.com/en/pepper}, QT\footnote{https://luxai.com/}, Nao\footnote{https://corporate-internal-prod.aldebaran.com/en/nao}, and Furhat\footnote{https://www.furhatrobotics.com/} as seen in Figure \ref{fig4}. Firstly students discussed the different \textbf{existing robots' appearances and capabilities}. Specifically how design choices, such as form, facial features, and size, shaped their perceptions of the robots' potential roles. QT was described as resembling an astronaut, with an endearing and friendly appearance. This friendliness made it particularly appealing for roles such as venting partner and study companion. The size and positioning of the robot were noted as important factors; participants imagined QT being effective at eye level on a desk, where it could create a more conversational dynamic. Furhat’s customisable digital face was seen as having potential for social communication practice. Students appreciated that its face could be programmed to show a range of facial expressions, and to simulate emotions and enable more realistic social interactions. For example, the idea of personalising the face to resemble someone familiar, like a director of studies, was seen as useful for practising conversations. Pepper’s design, with an integrated screen, was seen as advantageous for roles like presentation rehearsal partner and signposting. The screen offered functionality to display cues or additional information during interactions, making it useful for combining multiple tasks, such as presenting while receiving feedback or guidance. Nao was perceived as a more serious and businesslike robot, making it potentially suitable for structured roles like advocacy practice or study companion. 

Another theme related to designing the robot role was \textbf{affordances}. Students agreed that customisable voices could enhance the versatility of robots across different roles, a venting partner might require a voice that could range from serious to approachable, depending on the context and emotional needs of the user: "The venting partner… you might want them to sound more serious or more approachable, depending on what you're venting about." Similarly, advocacy practice and signposting roles might benefit from a more formal, serious tone to reflect confidence and authority. However, voices used for study companions or social communication practice should prioritise approachability and warmth. Movement was another critical factor in robot design, especially for roles like signposting and social communication practice. Participants noted that mobility could help robots guide users through physical spaces or point out locations: "If you go into a building and they say somewhere in this building there is the student support service, it might be helpful to have something that moves." On the other hand, movement was considered less relevant for roles like venting partners, where the focus is on emotional engagement rather than physical navigation. Participants discussed how the robot's size should align with its intended purpose and environment. Smaller robots were seen as more suitable for personal roles, such as venting partners or study companions, due to their non-threatening and approachable nature: "Personal robots would be smaller… if it was the same size as me, it would be a bit threatening." Conversely, larger robots were seen as more appropriate for community-oriented roles like signposting, where visibility and presence in public spaces were essential. 

\subsubsection{Robot Characteristics Design}
The second part of focus group 2 was related to designing characteristics of the robot and the first theme concerned \textbf{identity}. A noteworthy preference was expressed for robots that avoid explicit identity markers, such as age or gender. Neutrality was valued for its ability to sidestep biases, stereotypes, reducing the potential for prejudice and making the robot a distinct entity, separate from human social categories: "I like the robot-looking ones because they’re just robots. It’s a kind of blandness, but not in a way that feels empty—it’s a separate entity, which I like." Customisability emerged as a recurring theme, and participants emphasised the need for robots to adapt to individual user preferences and identities: "There might be some people who would want their robot to be as much like them… but there might be other people that want robots to have a completely separate identity." Participants noted the influence of gender in robot design and raised concerns about reinforcing stereotypes, particularly with the prevalence of female-presenting AI systems: "There are too many female AI assistants—it just feels icky. Why is it always a woman? Patriarchy can help me instead." Some students pointed out challenges of achieving true gender neutrality, and how societal perceptions shape even ostensibly androgynous designs: "What we consider gender-neutral often skews masculine, which is strange. Androgyny is technically vague, but in practice, it’s tied to specific characteristics." However, there were also reflections on the potential benefits of gendered robots in certain contexts. For example, some users might feel more comfortable or find the robot more approachable if its gender aligns with their own or conveys certain traits: "If you identify with the same gender the robot presents, it might feel more approachable, like a friend." Others noted how gendered traits could influence perceptions of competence or trustworthiness: "I would trust a female-presenting robot to signpost better because she seems like she knows what she’s doing." Students discussed how perceptions of age could influence interactions. Robots resembling children, for instance, were seen as potentially limiting for certain roles, such as venting partners: "I wouldn’t want to vent to it if it looked like a small child. It feels like yelling at a toddler." Conversely, robots expressing wisdom or maturity were seen as better suited for providing guidance or support. Rather than embodying a specific age, participants preferred designs that conveyed knowledge and understanding through their communication style: "I’d like it to feel wise in the way it communicates, not tied to a particular age."

\textbf{Language and voice} was another important consideration and students discussed the importance of robots supporting multiple languages and enabling smooth transitions between them: "I tend to speak a lot of Danglish [Danish + English]. It would be nice to have the option to switch languages easily, like changing the language on a phone." The ability to practice or use one’s native language was especially valued: "I usually speak French with ChatGPT—it’s good for practising. If my first language were different, I’d definitely want that option." Beyond simple multilingualism, participants expressed a desire for robots to handle code-switching, switching between languages or linguistic contexts (e.g., formal vs. informal): "It would be really cool if it could [code switch] depending on the combinations of languages or the combinations of different contexts within a language." Participants noted that this would require the robot to adapt to personal language usage over time, making it feel like a personalised companion: "If it could respond to a configured person, learning how you use language, that would be really cool." Adapting the tone and level of formality based on the social or contextual setting was another important feature to enhance the robot's ability to function as a trusted and versatile communication partner: "It would be great if it could switch between formal and informal depending on the situation or person."

The theme of robot \textbf{personality} showed a strong emphasis on adaptability and aligning a robot's personality with the specific context or task at hand. For example, in more formal settings like signposting, they preferred a personality that was reserved, professional, and even accompanied by a voice that felt appropriate for serious tasks. In more casual interactions, a more expressive, engaging and personified personality was seen as desirable. Participants noted that the personality of a robot could significantly affect its role as a study companion. While some users found it motivating to have a supportive and encouraging presence that cheered them on, others preferred a more disciplined personality, reminding them to stay focused and avoid distractions and being "brutally honest". Some mentioned the value of a personality that could support active engagement, such as one that would listen and respond as students explained concepts aloud to aid their understanding. At the same time, they expressed the need for the robot to adopt a quieter, more reserved role during moments of intense focus, such as by muting itself or playing music. A recurring theme was the robot’s potential to provide feedback and a strong preference for constructive criticism, prioritising honesty over emotional cushioning. However, they also acknowledged the need for nuanced feedback, and the ability to “dial up or down” the robot’s level of critique was seen as an important feature.

\subsubsection{Robot Contextual Design}
The final part of focus group 2 was related to contextual design dimensions of the robot. Choosing to discuss these aspects of design with the students was based on work previously done \cite{axelsson2021social, ostrowski2022ethics, markelius_design_justice, rakova2023terms} to extend co-design beyond mere consultation of superficial, robot-based dimensions, into dimensions that also affect the interaction, but is often overlooked. For example, \cite{rakova2023terms} present a framework to allow co-design of terms of service and user agreements related to algorithmic systems to empower the intended users to have meaningful impact on the conditions under which they use the technology. They also highlight the importance of consent, and how consent can be discussed in a co-design context to allow participants to help design the way in which meaningful consent can be given, and withdrawn, to use the technology. 

Hence, the first theme of contextual design concerns \textbf{consent and user agreements}. Many participants expressed concerns about the impracticality of traditional, lengthy online terms and conditions, and that expecting users to engage meaningfully with 50-page agreements is unreasonable. They suggested that consent must be reimagined for robotic systems, particularly in scenarios where interaction is immediate or community-based. For example, with public-use robots designed for tasks like signposting, participants believed a less detailed user agreement would suffice, allowing interactions to proceed quickly without requiring extensive individual consent at the point of use, and that they might rely on simplified consent protocols or generalised community-level agreements to avoid functionality disruptions. In contrast, personal robots, designed to engage in more sensitive, individualised tasks such as reading body language or storing personal information, were seen to require far more detailed terms and conditions. Layered consent structures were suggested in such cases, so that students can engage with summarised, conversationally delivered agreements while retaining access to a comprehensive, full-length version for those who wished to review the details in depth. For instance, one participant suggested that a robot might provide a friendly, conversational summary of its practices—e.g., “Do you agree that I might store our conversation for three days in my memory banks, after which it will be permanently deleted?”—to enhance the clarity and accessibility of consent. This approach was seen as preferable to abruptly halting interaction to demand agreement with dense legal text, which might discourage meaningful engagement. Participants also noted the importance of giving users the opportunity to revisit and review the terms of their agreement after initial consent had been given. They said that while many users might be satisfied with simplified summaries, the availability of the full terms and conditions should be preserved for those who wanted greater transparency. One participant remarked on the need for "constructive language" in these agreements, so that users understood the implications of their consent without feeling overwhelmed or misled. 

Finally, the theme of \textbf{data and privacy} elicited reflections on balancing trust, transparency, and usability, participants expressed strong concerns about the ethical handling of data, and the need for robust safeguards before they would feel comfortable engaging with a social robot. Many participants had a preference for systems that operate locally, such as on a personal device, as this was seen as inherently more secure and less prone to misuse by external entities. Ethical behaviour by companies was highlighted as a critical factor, with participants expressing a desire for clear evidence that organisations were adhering to high standards of data security and integrity. A key point of discussion centred on the importance of giving users options when it came to data collection. Participants suggested that robots should offer alternative forms of interaction for those who did not feel comfortable sharing personal data. For instance, a robot could provide general, non-personalised assistance, such as directing users to a relevant website or physical location where they could receive help without the need for data collection. This approach was seen as a way to empower users while still making the robot a valuable resource. The role of robots in relation to human workers was also discussed in the context of privacy. Participants generally agreed that robots were unlikely to fully replace human roles, particularly in tasks such as signposting or providing assistance. Instead, robots were seen as tools that could streamline processes, make interactions more approachable, or complement human efforts. This approach, where users could always turn to a human if they preferred not to engage with the robot, was seen as a way to alleviate data concerns while still maintaining accessibility. Participants suggested that robots could explicitly inform users of these alternatives, saying, for example, “If you do not consent to me collecting or storing this data, you can visit the help desk down the corridor.” 

\subsection{Summary Phase 2}
In phase 2, we conducted two focus groups with disabled students. The first focus group was centred around understanding the problem space, and ideating social robots for disability. For current technology usage, most participants relied on tools like screen readers and dictation software to enhance focus and access to learning materials, with audio-visual integration aiding comprehension. However, issues were mentioned like unnatural voice cadences, lack of multilingual support, and unclear academic guidelines for chatbot use. Students valued proactive reassurance provided by centralised services. However, inconsistencies across departments and the socially demanding nature of interactions with university staff created systematic frictions. When envisaging social robots, students saw them performing specific, low-energy tasks, such as assisting with communication, time management, or creative expression. Desirable features included accurate personalisation, adaptability, and approachable designs that avoided human likeness. Continuous user feedback was emphasised as crucial for robots to remain responsive to evolving disability related needs and identities. Participants also raised concerns about intersectional biases in robot design and functionality, particularly regarding marginalisation and neurodiversity. 

The second focus group, held a few weeks after the first, included four of the original five participants. This session was centred around co-design, with a focus on envisioning specific aspects of a social robot to support mediation and advocacy for disabled students. The first part of the session centred on designing the robot's role. Six potential roles were identified from the ideas discussed in the first focus group: venting partner, social communication practice, study companion, signposting, presentation rehearsal partner, and advocacy practice. The appearance and form of each robot were found to have a noteworthy influence on participants' perceptions of its suitability for specific roles. Affordances such as voice customisability, movement, personality, and size emerged as critical design considerations. In the discussion robots' characteristics students expressed a preference for robots that avoided explicit identity markers, such as age or gender, valuing neutrality as a way to sidestep biases and stereotypes. Language and voice design were also crucial such as the robots supporting multiple languages and enabling seamless code-switching between linguistic contexts. Another recurring theme was the robot's personality. The final part of the session focused on contextual design, specifically exploring how factors like consent and data privacy could be integrated into the robot’s design. Participants proposed conversationally delivered summaries of agreements, allowing users to engage meaningfully with key points while maintaining access to full-length versions for greater transparency and privacy.

\section{Discussion}
We set out to explore the perspectives of disabled students and professionals with extensive experience of working closely with disabled student on the potential role of social robots in supporting students' self-advocacy. More specifically, in Phase 1 of the study we investigated what strategies disabled students currently employ to support communicating their needs and accessing adjustments and how disabled students currently engage with and navigate existing technologies, and how these technologies shape their self-advocacy and access to learning. In Phase 2, we focused on how disabled students envision and perceive the functional, relational, and affective roles of social robots in supporting mediation and advocacy, as well as what challenges, ethical concerns, and socio-technical assumptions disabled students identify in their envisions of social robots, and how these experiences reflect broader systemic exclusions and inequities.

The concerns and priorities around student advocacy and communication support identified in the results reflect previous findings insofar as issues around accessing agreed adjustments and encountering negative attitudes emerge strongly from the participants' experiences at the University of Cambridge.  Within the specific context of this institution, disabled students find the collegiate structure both a benefit (college as a valued additional source of support) and a challenge in adding to the complexity of the institutional structure. Similarly, the perceptions of the potential role of social robots reflect the affordances of this unique organisational structure, with aspects of the broad wellbeing role of college evident in the identification of small signposting, informational and guidance tasks as typically envisioned in social robot interactions. 

Whilst the unique features of the University of Cambridge are evident in much of the framing of the experiences discussed by the participants, the translation of these findings to the broader higher educational context requires only minor re-framing and generalising of the support roles and activities. The roles of disability practitioners and the legislative context  of reasonable adjustments are common across all UK universities. Therefore, the findings translate easily to other higher education contexts. Firstly, we discuss findings around diversity, difference and disability. We then discuss findings around co-design, communication and context before considering the limitations of this study and recommendations for next steps.

\subsection{Diversity, Difference and Disability}
The Social Model of Disability (SMD) \cite{oliver2013social} is implicit (and sometimes explicit) in the relationship between disabled students and their access to learning, also defining the potential for social robot support in terms of challenges as well as benefits.  The disabling features of the environment for disabled students are diverse and frequently encountered throughout the students' university experience. Whilst there are experiences of encounters with unsympathetic staff and systems, disabled students more frequently attribute challenges in communicating or accessing required adjustment to the environment (including lack of training, etc.) rather than a fault or deliberate act on the behalf of the individual staff member. By extension, the capacity of a social robot to effectively support advocacy is determined by the quality of the inputted data or programming. 
Furthermore, there are limits to the know-ability of the students' needs and preferences as even the most up to date information and language used to inform responses of a social robot wold still need to be aligned with the students' individual preferences (and change as these change).  Therefore, for disabled students, the key to technology that supports their advocacy is a high degree of control over the language and frames of reference (whether programmed or AI). Conversely, disabled students require a level of trust to build with the technology if factual information or communication suggestions can be accepted and employed by the students as part of their authentic interactions with staff and organisations within the HEI. Throughout these interactions, the risk of being misunderstood is ever present for disabled students \cite{heasman2018perspective}. Therefore, advocacy for disabled students, particularly neurodivergent students, requires understanding and recognition of the 'double empathy problem' \cite{Milton_2012}. 
An additional layer of complexity comes through the power dynamics inherent in the relationship between the disabled students and the HEI. Bourdieu's \cite{heffernan2022bourdieu} analysis of the entrenched marginalisation of disadvantaged groups extends into the potential use of social robots in student advocacy support, reflected in the depth of the concerns of disabled students over the ultimate control of the information shared within a social robot and whose interests (the student user or the HEI provider) are ultimately served by the advice or information being shared.  For disabled students, this marginalisation if often compounded by intersectionalities around identity (e.g. gender, sexuality, etc) or background, creating vulnerabilities across multiple dimensions to be recognised and addressed. 
Finally, some of the most pressing issues for disabled students engage with policy debates at the heart of current HEI policy, such as the availability of, and access to, lecture recordings \cite{nordmann2022lecture}. Therefore, spaces such as these where advocacy necessitates engagement with institutional policy rather than existing practice are seen by students and practitioners to present particular advocacy challenges that may be beyond the scope of technological interventions.  Thus, advocacy support (whether human, robot or other technology) must at some level engage with underlying dynamic complexities around diversity, difference and disability. 

\subsection{Co-Design, Communication and Context}
Our study adopted a co-design methodology, with the aim to enable a bottom-up approach to development and close collaboration with the communities and individuals for which the technology is intended. We involved disabled students directly to understand the problem space and for ideation, and ensured that their lived experiences and contextual knowledge shaped the design process. Scholars in social robotics have explored how ethics, equity and justice is enabled by research and design practices that align with frameworks such as design justice \cite{ostrowski2022ethics, costanza2020design}. Indeed, engaging disabled students and other stakeholders in co-design foregrounds their knowledge of systemic barriers and their specific needs, it shifts power dynamics by giving them greater control and influence over technologies that impact their lives. \cite{ostrowski2022ethics} prompts researchers in HRI to actively reflect on who is included in the design process from an contextual and intersectional point of view, i.e. considering race, class, gender, sexuality etc and not only focusing on e.g. disability per se. As shown in the findings of the current study, participants did indeed reflect on how their disabled experiences intersect with e.g. gender and race. Furthermore, the co-design process also aligned with other HRI equitable design questions \cite{ostrowski2022ethics} such as what values are prioritised in the design of the robot and/or robot application and what stories do people share about the robot and its impact on them. This approach enables communication of values and ideas through a medium that allows participants to speculate, reflect, envision and design freely, having a direct impact on the development of the social robot. Furthermore,the framework \textit{Terms-we-serve-with} \cite{rakova2023terms} further informed our approach to \textit{contextual design} extending the design beyond mere physical traits and superficial user preferences. For example, participants proposed conversationally delivered summaries of agreements to facilitate comprehension and engagement. Co-constitution foregrounds user agency, in the form of mutual assent and refusal mechanisms, empowering participants to opt out or customise their engagement with the system \cite{rakova2023terms}. Additionally, this also aligned with broader values of design justice \cite{costanza2020design}, enabling the creation of systems that deeply reflect contextual nuances, shying away from potentially harmful one-size-fits-all approaches. Future research in HRI, particularly aimed towards vulnerable populations should consider how to broaden the scope of participation, and adopt a critical lens on co-design to reflect on how it can empower individuals rather than being exploitative \cite{sloane2022participation, Bennett_Rosner_2019}. 

\subsection{Limitations}
The provision of disability support is particular to each HEI, so the findings discussed here are limited to the context of the University of Cambridge in terms of the sources and systems of support. In particular, the role of the students' college in advocacy and as a resource is unique to collegiate universities. Additionally, within the University of Cambridge, different departments have differing approaches to the regulation of AI. A further limitation of this study is the availability of and familiarity with emerging technologies for the participants. The data were collected in the first term of the 2024-25 academic year, reflecting common usage and knowledge of AI and other technologies at this time. 

\subsection{Recommendations for Future Research}
Based on the findings and discussion of this study, we offer a set of recommendations aimed at designers and developers of social robots for students with disabilities. These recommendations also extend to other stakeholders, such as roboticists working on social robots for other vulnerable populations, as well as policymakers and those involved in the governance of robotic technologies.

\begin{itemize}
    \item \textbf{Locating disability in the environment:} Adopt a social model of disability in the design and deployment of social robots as well as in framing of the overall research goals, objectives and research questions. This allows for shying away from the harmful notion that social robots are a "fix" to cure disabled people, into instead acknowledging the systemic and environmental barriers that hinder participation and inclusion, and the risk of social robots reproducing and further entrenching disabling features of the higher education environment. 
    \item \textbf{Neurodiversity framed empathy:} Incorporate the concept of the Double Empathy Problem in the design of social robots, which highlights the bidirectional challenges in understanding between neurodivergent and neurotypical individuals. What the notion of empathy means in the context of social robots for disabled people needs to be thoroughly investigated, and the risk for reproducing harmful, unauthentic empathy mitigated. 
    \item \textbf{Temporality:} Recognise the dynamic and evolving nature of disability, e.g. how individuals personally conceptualise their conditions over time, or societal discourse change. Social robots should reflect this shifting understanding of disability and interactions should enable temporal and continuous shaping of disability discourse as well as disabled peoples' experiences.
    \item \textbf{Anthropomorphism assumption:} Avoid the assumption that more human-likeness in appearance and behaviour is always desirable, as it may appear strange rather than helpful. In behaviour, decreased human-likeness could reduce risk for biases and prejudice stemming from human social categories. A less human-like appearance may draw from features found in animals, pets, or spherical, more robot-like designs that endow cuteness, friendliness and approachability. 
    \item \textbf{Contextual design:} Allow co-design of social robots, especially for vulnerable groups such as disabled students to extend beyond preferences of physical and non-physical traits into contextual design dimensions such as terms of service, inherent values and consent.
\end{itemize}

\section{Conclusion}
In this paper, we present the analysis and findings of an iterative, participatory study in two phases focusing on the design and usage of AI, social robots and LLMs for mediation and advocacy for students with disabilities in higher education. We conducted a bottom up approach, where we engaged with relevant stakeholders, such as disabled students, and professionals to understand the problem space and ideation of social robots as support for disabled students. Additionally, we engaged in co-design with the participants, and our findings show how disabled students envisage both physical, non-physical and contextual traits of social robots. We conducted thematic analyses of both interviews, as well as focus groups, and presented findings of potential roles, characteristics as well as limitations and ethical considerations of for such robots. Additionally, we discussed aspects of our methodology, as well as framing of disability in social robotic research and how it should be considered in the future. Finally we present a set of recommendations in order to help future roboticists, technologists, disability practitioners and policy-makers developing and deploying social robots, AI and other technologies to support disabled students in higher education. The findings from this study support the further exploration of social robots as a tools for supporting the medication and self-advocacy for disabled students, but emphasise the importance of reflecting the needs and concerns of disabled students and responding to the dynamic higher education disability context.

\begin{acks}
This work was supported by the  University of Cambridge School of Technology Seed Fund. Alva Markelius is supported by the Cambridge International Trust Scholarship. 
\end{acks}

\bibliographystyle{ACM-Reference-Format}
\bibliography{sample-base}


\begin{thebibliography}{68}


\ifx \showCODEN    \undefined \def \showCODEN     #1{\unskip}     \fi
\ifx \showDOI      \undefined \def \showDOI       #1{#1}\fi
\ifx \showISBNx    \undefined \def \showISBNx     #1{\unskip}     \fi
\ifx \showISBNxiii \undefined \def \showISBNxiii  #1{\unskip}     \fi
\ifx \showISSN     \undefined \def \showISSN      #1{\unskip}     \fi
\ifx \showLCCN     \undefined \def \showLCCN      #1{\unskip}     \fi
\ifx \shownote     \undefined \def \shownote      #1{#1}          \fi
\ifx \showarticletitle \undefined \def \showarticletitle #1{#1}   \fi
\ifx \showURL      \undefined \def \showURL       {\relax}        \fi
\providecommand\bibfield[2]{#2}
\providecommand\bibinfo[2]{#2}
\providecommand\natexlab[1]{#1}
\providecommand\showeprint[2][]{arXiv:#2}

\bibitem[{Accessibility \& Disability Resource Centre, University of Cambridge}(2025)]%
        {ADRC2025}
\bibfield{author}{\bibinfo{person}{{Accessibility \& Disability Resource Centre, University of Cambridge}}.} \bibinfo{year}{2025}\natexlab{}.
\newblock \bibinfo{title}{Accessibility \& Disability Resource Centre (ADRC): Annual Report 2023-2024}.
\newblock
\newblock
\urldef\tempurl%
\url{https://www.disability.admin.cam.ac.uk/sites/default/files/2025-02/ADRC%20Annual%20Report%2023-24%20Accessible%20FINAL.pdf}
\showURL{%
\tempurl}
\newblock
\shownote{Accessed: 2025-02-18}.


\bibitem[Axelsson et~al\mbox{.}(2021)]%
        {axelsson2021social}
\bibfield{author}{\bibinfo{person}{Minja Axelsson}, \bibinfo{person}{Raquel Oliveira}, \bibinfo{person}{Mattia Racca}, {and} \bibinfo{person}{Ville Kyrki}.} \bibinfo{year}{2021}\natexlab{}.
\newblock \showarticletitle{Social robot co-design canvases: A participatory design framework}.
\newblock \bibinfo{journal}{\emph{ACM Transactions on Human-Robot Interaction (THRI)}} \bibinfo{volume}{11}, \bibinfo{number}{1} (\bibinfo{year}{2021}), \bibinfo{pages}{1--39}.
\newblock
\urldef\tempurl%
\url{https://doi.org/10.1145/3472225}
\showDOI{\tempurl}


\bibitem[Axelsson et~al\mbox{.}(2024)]%
        {axelsson2024robots}
\bibfield{author}{\bibinfo{person}{Minja Axelsson}, \bibinfo{person}{Micol Spitale}, {and} \bibinfo{person}{Hatice Gunes}.} \bibinfo{year}{2024}\natexlab{}.
\newblock \showarticletitle{Robots as mental well-being coaches: Design and ethical recommendations}.
\newblock \bibinfo{journal}{\emph{ACM Transactions on Human-Robot Interaction}} \bibinfo{volume}{13}, \bibinfo{number}{2} (\bibinfo{year}{2024}), \bibinfo{pages}{1--55}.
\newblock
\urldef\tempurl%
\url{https://doi.org/10.1145/3643457}
\showDOI{\tempurl}


\bibitem[Bailey(2024)]%
        {bailey2024neurodiversity}
\bibfield{author}{\bibinfo{person}{Julie Bailey}.} \bibinfo{year}{2024}\natexlab{}.
\newblock \emph{\bibinfo{title}{Neurodiversity and Learning Engagement in Higher Education}}.
\newblock \bibinfo{thesistype}{Ph.\,D. Dissertation}. \bibinfo{school}{University of Cambridge}.
\newblock


\bibitem[Baron-Cohen(2017)]%
        {baron2017editorial}
\bibfield{author}{\bibinfo{person}{Simon Baron-Cohen}.} \bibinfo{year}{2017}\natexlab{}.
\newblock \bibinfo{title}{Editorial Perspective: Neurodiversity--a revolutionary concept for autism and psychiatry}.
\newblock , \bibinfo{numpages}{744--747}~pages.
\newblock


\bibitem[Bennett and Rosner(2019)]%
        {Bennett_Rosner_2019}
\bibfield{author}{\bibinfo{person}{Cynthia~L. Bennett} {and} \bibinfo{person}{Daniela~K. Rosner}.} \bibinfo{year}{2019}\natexlab{}.
\newblock \showarticletitle{The Promise of Empathy: Design, Disability, and Knowing the “Other”}. In \bibinfo{booktitle}{\emph{Proceedings of the 2019 CHI Conference on Human Factors in Computing Systems}} \emph{(\bibinfo{series}{CHI ’19})}. \bibinfo{publisher}{Association for Computing Machinery}, \bibinfo{address}{New York, NY, USA}, \bibinfo{pages}{1–13}.
\newblock
\showISBNx{978-1-4503-5970-2}
\urldef\tempurl%
\url{https://doi.org/10.1145/3290605.3300528}
\showDOI{\tempurl}


\bibitem[Bertacchini et~al\mbox{.}(2023)]%
        {Bertacchini_Demarco_Scuro_Pantano_Bilotta_2023}
\bibfield{author}{\bibinfo{person}{Francesca Bertacchini}, \bibinfo{person}{Francesco Demarco}, \bibinfo{person}{Carmelo Scuro}, \bibinfo{person}{Pietro Pantano}, {and} \bibinfo{person}{Eleonora Bilotta}.} \bibinfo{year}{2023}\natexlab{}.
\newblock \showarticletitle{A social robot connected with chatGPT to improve cognitive functioning in ASD subjects}.
\newblock \bibinfo{journal}{\emph{Frontiers in Psychology}}  \bibinfo{volume}{14} (\bibinfo{year}{2023}), \bibinfo{pages}{01--22}.
\newblock
\showISSN{1664-1078}
\urldef\tempurl%
\url{https://doi.org/10.3389/fpsyg.2023.1232177}
\showDOI{\tempurl}


\bibitem[Braun and Clarke(2006)]%
        {braun2006using}
\bibfield{author}{\bibinfo{person}{Virginia Braun} {and} \bibinfo{person}{Victoria Clarke}.} \bibinfo{year}{2006}\natexlab{}.
\newblock \showarticletitle{Using thematic analysis in psychology}.
\newblock \bibinfo{journal}{\emph{Qualitative research in psychology}} \bibinfo{volume}{3}, \bibinfo{number}{2} (\bibinfo{year}{2006}), \bibinfo{pages}{77--101}.
\newblock
\urldef\tempurl%
\url{https://doi.org/10.1191/1478088706qp063oa}
\showDOI{\tempurl}


\bibitem[Brewer et~al\mbox{.}(2023)]%
        {brewer2023disabled}
\bibfield{author}{\bibinfo{person}{Gayle Brewer}, \bibinfo{person}{Emily Urwin}, {and} \bibinfo{person}{Beth Witham}.} \bibinfo{year}{2023}\natexlab{}.
\newblock \showarticletitle{Disabled student experiences of Higher Education}.
\newblock \bibinfo{journal}{\emph{Disability \& Society}} \bibinfo{volume}{40}, \bibinfo{number}{1} (\bibinfo{year}{2023}), \bibinfo{pages}{108--127}.
\newblock
\urldef\tempurl%
\url{https://doi.org/10.1080/09687599.2023.2263633}
\showDOI{\tempurl}


\bibitem[Bruce and Aylward(2021)]%
        {bruce2021disability}
\bibfield{author}{\bibinfo{person}{Cynthia Bruce} {and} \bibinfo{person}{M~Lynn Aylward}.} \bibinfo{year}{2021}\natexlab{}.
\newblock \showarticletitle{Disability and Self-Advocacy Experiences in University Learning Contexts.}
\newblock \bibinfo{journal}{\emph{Scandinavian Journal of Disability Research}} \bibinfo{volume}{23}, \bibinfo{number}{1} (\bibinfo{year}{2021}), \bibinfo{pages}{14–26}.
\newblock
\urldef\tempurl%
\url{https://doi.org/10.16993/sjdr.741}
\showDOI{\tempurl}


\bibitem[Costanza-Chock(2020)]%
        {costanza2020design}
\bibfield{author}{\bibinfo{person}{Sasha Costanza-Chock}.} \bibinfo{year}{2020}\natexlab{}.
\newblock \bibinfo{booktitle}{\emph{Design justice: Community-led practices to build the worlds we need}}.
\newblock \bibinfo{publisher}{The MIT Press}, \bibinfo{address}{Cambridge, Massachusetts}.
\newblock


\bibitem[Dam et~al\mbox{.}(2022)]%
        {dam_experiences_2022}
\bibfield{author}{\bibinfo{person}{Kirstin~van Dam}, \bibinfo{person}{Marieke Gielissen}, \bibinfo{person}{Rachelle Reijnders}, \bibinfo{person}{Agnes van~der Poel}, {and} \bibinfo{person}{Brigitte Boon}.} \bibinfo{year}{2022}\natexlab{}.
\newblock \showarticletitle{Experiences of {Persons} {With} {Executive} {Dysfunction} in {Disability} {Care} {Using} a {Social} {Robot} to {Execute} {Daily} {Tasks} and {Increase} the {Feeling} of {Independence}: {Multiple}-{Case} {Study}}.
\newblock \bibinfo{journal}{\emph{JMIR Rehabilitation and Assistive Technologies}} \bibinfo{volume}{9}, \bibinfo{number}{4} (\bibinfo{date}{Nov.} \bibinfo{year}{2022}), \bibinfo{pages}{e41313}.
\newblock
\urldef\tempurl%
\url{https://doi.org/10.2196/41313}
\showDOI{\tempurl}
\newblock
\shownote{Company: JMIR Rehabilitation and Assistive Technologies Distributor: JMIR Rehabilitation and Assistive Technologies Institution: JMIR Rehabilitation and Assistive Technologies Label: JMIR Rehabilitation and Assistive Technologies Publisher: JMIR Publications Inc., Toronto, Canada}.


\bibitem[Dehnert(2024)]%
        {dehnert2024ability}
\bibfield{author}{\bibinfo{person}{Marco Dehnert}.} \bibinfo{year}{2024}\natexlab{}.
\newblock \showarticletitle{Ability and Disability: Social Robots and Accessibility, Disability Justice, and the Socially Constructed Normal Body}.
\newblock \bibinfo{journal}{\emph{The De Gruyter Handbook of Robots in Society and Culture}}  \bibinfo{volume}{3} (\bibinfo{year}{2024}), \bibinfo{pages}{429}.
\newblock


\bibitem[Di~Miceli(2024)]%
        {di2024diversity}
\bibfield{author}{\bibinfo{person}{Mathieu Di~Miceli}.} \bibinfo{year}{2024}\natexlab{}.
\newblock \showarticletitle{Diversity in the United Kingdom: Quantification for higher education in comparison to the general population}.
\newblock \bibinfo{journal}{\emph{European Journal of Education}} \bibinfo{volume}{59}, \bibinfo{number}{2} (\bibinfo{year}{2024}), \bibinfo{pages}{e12595}.
\newblock


\bibitem[{Disabled Students UK}(2024)]%
        {disabledstudentsuk2024}
\bibfield{author}{\bibinfo{person}{{Disabled Students UK}}.} \bibinfo{year}{2024}\natexlab{}.
\newblock \bibinfo{title}{Access Insights Report 2024: Framework and Baseline}.
\newblock
\newblock
\urldef\tempurl%
\url{http://accessinsights.co.uk/}
\showURL{%
\tempurl}
\newblock
\shownote{Available at Disabled Students UK website.}.


\bibitem[{Equality and Human Rights Commission}(2024)]%
        {EHRC2024}
\bibfield{author}{\bibinfo{person}{{Equality and Human Rights Commission}}.} \bibinfo{year}{2024}\natexlab{}.
\newblock \bibinfo{title}{Advice note for the higher education sector from the legal case of University of Bristol vs Abrahart}.
\newblock
\newblock
\urldef\tempurl%
\url{https://www.equalityhumanrights.com/guidance/advice-note-higher-education-sector-legal-case-university-bristol-vs-abrahart}
\showURL{%
\tempurl}
\newblock
\shownote{Accessed: 2025-01-15}.


\bibitem[Fereday and Muir-Cochrane(2006)]%
        {fereday2006demonstrating}
\bibfield{author}{\bibinfo{person}{Jennifer Fereday} {and} \bibinfo{person}{Eimear Muir-Cochrane}.} \bibinfo{year}{2006}\natexlab{}.
\newblock \showarticletitle{Demonstrating rigor using thematic analysis: A hybrid approach of inductive and deductive coding and theme development}.
\newblock \bibinfo{journal}{\emph{International journal of qualitative methods}} \bibinfo{volume}{5}, \bibinfo{number}{1} (\bibinfo{year}{2006}), \bibinfo{pages}{80--92}.
\newblock
\urldef\tempurl%
\url{https://doi.org/10.1177/160940690600500107}
\showDOI{\tempurl}


\bibitem[Fiora et~al\mbox{.}(2024)]%
        {Fiora_Piferi_Crovari_Garzotto_2024}
\bibfield{author}{\bibinfo{person}{Asterio Fiora}, \bibinfo{person}{Francesco Piferi}, \bibinfo{person}{Pietro Crovari}, {and} \bibinfo{person}{Franca Garzotto}.} \bibinfo{year}{2024}\natexlab{}.
\newblock \showarticletitle{Exploring Large Language Models for the Education of Individuals with Cognitive Impairments}. In \bibinfo{booktitle}{\emph{INTED2024 Proceedings}}. \bibinfo{publisher}{18th International Technology, Education and Development Conference}, \bibinfo{address}{Valencia, Spain}, \bibinfo{pages}{4479–4487}.
\newblock
\urldef\tempurl%
\url{https://doi.org/10.21125/inted.2024.1161}
\showDOI{\tempurl}


\bibitem[Francis(2015)]%
        {Francis_2015}
\bibfield{author}{\bibinfo{person}{Leslie~P. Francis}.} \bibinfo{year}{2015}\natexlab{}.
\newblock \showarticletitle{Disability and Philosophy: Applying Ethics in Circumstances of Injustice}.
\newblock \bibinfo{journal}{\emph{Social Science Research Network}} \bibinfo{volume}{42}, \bibinfo{number}{173} (\bibinfo{date}{Dec.} \bibinfo{year}{2015}), \bibinfo{pages}{35--36}.
\newblock
\urldef\tempurl%
\url{https://doi.org/10.2139/ssrn.2791661}
\showDOI{\tempurl}


\bibitem[Frennert et~al\mbox{.}(2024)]%
        {Frennert_Persson_Skavron_2024}
\bibfield{author}{\bibinfo{person}{Susanne Frennert}, \bibinfo{person}{Johanna Persson}, {and} \bibinfo{person}{Sarah Skavron}.} \bibinfo{year}{2024}\natexlab{}.
\newblock \showarticletitle{A Critical Narrative Review of Assistive Robotics and Call for a Systems and User-Centered Approaches to Enhance Quality of Life of Individuals with Disabilities}. In \bibinfo{booktitle}{\emph{Adjunct Proceedings of the 2024 Nordic Conference on Human-Computer Interaction}}. \bibinfo{publisher}{ACM}, \bibinfo{address}{Uppsala Sweden}, \bibinfo{pages}{1–11}.
\newblock
\showISBNx{9798400709654}
\urldef\tempurl%
\url{https://doi.org/10.1145/3677045.3685495}
\showDOI{\tempurl}


\bibitem[Gadiraju et~al\mbox{.}(2023)]%
        {Gadiraju_Kane_Dev_Taylor_Wang_Denton_Brewer_2023}
\bibfield{author}{\bibinfo{person}{Vinitha Gadiraju}, \bibinfo{person}{Shaun Kane}, \bibinfo{person}{Sunipa Dev}, \bibinfo{person}{Alex Taylor}, \bibinfo{person}{Ding Wang}, \bibinfo{person}{Emily Denton}, {and} \bibinfo{person}{Robin Brewer}.} \bibinfo{year}{2023}\natexlab{}.
\newblock \showarticletitle{“I wouldn’t say offensive but...”: Disability-Centered Perspectives on Large Language Models}. In \bibinfo{booktitle}{\emph{2023 ACM Conference on Fairness, Accountability, and Transparency}}. \bibinfo{publisher}{ACM}, \bibinfo{address}{Chicago IL USA}, \bibinfo{pages}{205–216}.
\newblock
\showISBNx{9798400701924}
\urldef\tempurl%
\url{https://doi.org/10.1145/3593013.3593989}
\showDOI{\tempurl}


\bibitem[Garcia-Pi et~al\mbox{.}(2023)]%
        {Garcia-Pi_Chaudhury_Versaw_Back_Kwon_Kicklighter_Taele_Seo_2023}
\bibfield{author}{\bibinfo{person}{Brittany Garcia-Pi}, \bibinfo{person}{Rohan Chaudhury}, \bibinfo{person}{Miles Versaw}, \bibinfo{person}{Jonathan Back}, \bibinfo{person}{Dongjin Kwon}, \bibinfo{person}{Caleb Kicklighter}, \bibinfo{person}{Paul Taele}, {and} \bibinfo{person}{Jinsil~Hwaryoung Seo}.} \bibinfo{year}{2023}\natexlab{}.
\newblock \showarticletitle{AllyChat: Developing a VR Conversational AI Agent Using Few-Shot Learning to Support Individuals with Intellectual Disabilities}. In \bibinfo{booktitle}{\emph{Human-Computer Interaction – INTERACT 2023}}, \bibfield{editor}{\bibinfo{person}{José Abdelnour~Nocera}, \bibinfo{person}{Marta Kristín~Lárusdóttir}, \bibinfo{person}{Helen Petrie}, \bibinfo{person}{Antonio Piccinno}, {and} \bibinfo{person}{Marco Winckler}} (Eds.). \bibinfo{publisher}{Springer Nature Switzerland}, \bibinfo{address}{Cham}, \bibinfo{pages}{402–407}.
\newblock
\showISBNx{978-3-031-42293-5}
\urldef\tempurl%
\url{https://doi.org/10.1007/978-3-031-42293-5_43}
\showDOI{\tempurl}


\bibitem[Goering(2015)]%
        {Goering_2015}
\bibfield{author}{\bibinfo{person}{Sara Goering}.} \bibinfo{year}{2015}\natexlab{}.
\newblock \showarticletitle{Rethinking disability: the social model of disability and chronic disease}.
\newblock \bibinfo{journal}{\emph{Current Reviews in Musculoskeletal Medicine}} \bibinfo{volume}{8}, \bibinfo{number}{2} (\bibinfo{date}{April} \bibinfo{year}{2015}), \bibinfo{pages}{134}.
\newblock
\urldef\tempurl%
\url{https://doi.org/10.1007/s12178-015-9273-z}
\showDOI{\tempurl}


\bibitem[Government(2010)]%
        {uk_equality_act_2010}
\bibfield{author}{\bibinfo{person}{UK Government}.} \bibinfo{year}{2010}\natexlab{}.
\newblock \bibinfo{title}{Equality Act 2010}.
\newblock
\newblock
\newblock
\shownote{Available at: \url{https://www.legislation.gov.uk/ukpga/2010/15/contents} [Accessed November 26, 2024]}.


\bibitem[{Gov.uk}(nd)]%
        {govuk_dsa}
\bibfield{author}{\bibinfo{person}{{Gov.uk}}.} \bibinfo{year}{n.d.}\natexlab{}.
\newblock \bibinfo{title}{Disabled Students' Allowance (DSA)}.
\newblock \bibinfo{howpublished}{\url{https://www.gov.uk/disabled-students-allowance-dsa}}.
\newblock
\newblock
\shownote{Accessed: 2025-01-21}.


\bibitem[Guerreiro et~al\mbox{.}(2019)]%
        {guerreiro_cabot_2019}
\bibfield{author}{\bibinfo{person}{João Guerreiro}, \bibinfo{person}{Daisuke Sato}, \bibinfo{person}{Saki Asakawa}, \bibinfo{person}{Huixu Dong}, \bibinfo{person}{Kris~M. Kitani}, {and} \bibinfo{person}{Chieko Asakawa}.} \bibinfo{year}{2019}\natexlab{}.
\newblock \showarticletitle{{CaBot}: {Designing} and {Evaluating} an {Autonomous} {Navigation} {Robot} for {Blind} {People}}. In \bibinfo{booktitle}{\emph{Proceedings of the 21st {International} {ACM} {SIGACCESS} {Conference} on {Computers} and {Accessibility}}} \emph{(\bibinfo{series}{{ASSETS} '19})}. \bibinfo{publisher}{Association for Computing Machinery}, \bibinfo{address}{New York, NY, USA}, \bibinfo{pages}{68--82}.
\newblock
\showISBNx{978-1-4503-6676-2}
\urldef\tempurl%
\url{https://doi.org/10.1145/3308561.3353771}
\showDOI{\tempurl}


\bibitem[Haynes and Loblay(2024)]%
        {haynes2024rethinking}
\bibfield{author}{\bibinfo{person}{Abby Haynes} {and} \bibinfo{person}{Victoria Loblay}.} \bibinfo{year}{2024}\natexlab{}.
\newblock \showarticletitle{Rethinking Barriers and Enablers in Qualitative Health Research: Limitations, Alternatives, and Enhancements}.
\newblock \bibinfo{journal}{\emph{Qualitative Health Research}} \bibinfo{volume}{34}, \bibinfo{number}{14} (\bibinfo{year}{2024}), \bibinfo{pages}{1371--1383}.
\newblock


\bibitem[Heasman and Gillespie(2018)]%
        {heasman2018perspective}
\bibfield{author}{\bibinfo{person}{Brett Heasman} {and} \bibinfo{person}{Alex Gillespie}.} \bibinfo{year}{2018}\natexlab{}.
\newblock \showarticletitle{Perspective-taking is two-sided: Misunderstandings between people with Asperger’s syndrome and their family members}.
\newblock \bibinfo{journal}{\emph{Autism}} \bibinfo{volume}{22}, \bibinfo{number}{6} (\bibinfo{year}{2018}), \bibinfo{pages}{740--750}.
\newblock


\bibitem[Heffernan(2022)]%
        {heffernan2022bourdieu}
\bibfield{author}{\bibinfo{person}{Troy Heffernan}.} \bibinfo{year}{2022}\natexlab{}.
\newblock \bibinfo{booktitle}{\emph{Bourdieu and higher education: Life in the modern university}}.
\newblock \bibinfo{publisher}{Springer Nature}, \bibinfo{address}{Springer Singapore}.
\newblock


\bibitem[Horlin et~al\mbox{.}(2024)]%
        {horlin2024can}
\bibfield{author}{\bibinfo{person}{Chiara Horlin}, \bibinfo{person}{Barbora Hronska}, {and} \bibinfo{person}{Emily Nordmann}.} \bibinfo{year}{2024}\natexlab{}.
\newblock \showarticletitle{I can be a “normal” student: The role of lecture capture in supporting disabled and neurodivergent students’ participation in higher education}.
\newblock \bibinfo{journal}{\emph{Higher Education}}  \bibinfo{volume}{88} (\bibinfo{year}{2024}), \bibinfo{pages}{2075–2092}.
\newblock
\urldef\tempurl%
\url{https://doi.org/10.1007/s10734-024-01201-5}
\showDOI{\tempurl}


\bibitem[Howard and Sedgewick(2021)]%
        {howard2021anything}
\bibfield{author}{\bibinfo{person}{Philippa~L Howard} {and} \bibinfo{person}{Felicity Sedgewick}.} \bibinfo{year}{2021}\natexlab{}.
\newblock \showarticletitle{‘Anything but the phone!’: Communication mode preferences in the autism community}.
\newblock \bibinfo{journal}{\emph{Autism}} \bibinfo{volume}{25}, \bibinfo{number}{8} (\bibinfo{year}{2021}), \bibinfo{pages}{2265--2278}.
\newblock


\bibitem[Khasawneh(2024)]%
        {Khasawneh_2024}
\bibfield{author}{\bibinfo{person}{Mohamad Ahmad~Saleem Khasawneh}.} \bibinfo{year}{2024}\natexlab{}.
\newblock \showarticletitle{Teacher Opinions on the Role of Educational Robots in Enhancing Programming Skills among Hearing-Impaired Students}.
\newblock \bibinfo{journal}{\emph{International Journal of Learning, Teaching and Educational Research}} \bibinfo{volume}{23}, \bibinfo{number}{55} (\bibinfo{date}{May} \bibinfo{year}{2024}), \bibinfo{pages}{309–322}.
\newblock
\showISSN{1694-2116}
\urldef\tempurl%
\url{https://doi.org/10.26803/ijlter.23.5.16}
\showDOI{\tempurl}


\bibitem[Kian et~al\mbox{.}(2024)]%
        {Kian_Zong}
\bibfield{author}{\bibinfo{person}{Mina~J. Kian}, \bibinfo{person}{Mingyu Zong}, \bibinfo{person}{Katrin Fischer}, \bibinfo{person}{Abhyuday Singh}, \bibinfo{person}{Anna-Maria Velentza}, \bibinfo{person}{Pau Sang}, \bibinfo{person}{Shriya Upadhyay}, \bibinfo{person}{Anika Gupta}, \bibinfo{person}{Misha~A. Faruki}, \bibinfo{person}{Wallace Browning}, \bibinfo{person}{Sebastien M.~R. Arnold}, \bibinfo{person}{Bhaskar Krishnamachari}, {and} \bibinfo{person}{Maja~J. Mataric}.} \bibinfo{year}{2024}\natexlab{}.
\newblock \bibinfo{title}{Can an LLM-Powered Socially Assistive Robot Effectively and Safely Deliver Cognitive Behavioral Therapy? A Study With University Students}.  (\bibinfo{year}{2024}).
\newblock
\urldef\tempurl%
\url{https://doi.org/10.48550/arXiv.2402.17937}
\showDOI{\tempurl}
\newblock
\shownote{arXiv:2402.17937}.


\bibitem[Lalwani et~al\mbox{.}(2024)]%
        {lalwani2024productivity}
\bibfield{author}{\bibinfo{person}{Himanshi Lalwani}, \bibinfo{person}{Maha Elgarf}, {and} \bibinfo{person}{Hanan Salam}.} \bibinfo{year}{2024}\natexlab{}.
\newblock \bibinfo{title}{Productivity CoachBot: a Social Robot Coach for University Students with ADHD}.  (\bibinfo{year}{2024}).
\newblock
\newblock
\shownote{Preprint}.


\bibitem[Lister et~al\mbox{.}(2021)]%
        {lister_taylor_2021}
\bibfield{author}{\bibinfo{person}{Kate Lister}, \bibinfo{person}{Tim Coughlan}, \bibinfo{person}{Ian Kenny}, \bibinfo{person}{Ruth Tudor}, {and} \bibinfo{person}{Francisco Iniesto}.} \bibinfo{year}{2021}\natexlab{}.
\newblock \showarticletitle{Taylor, the {Disability} {Disclosure} {Virtual} {Assistant}: {A} {Case} {Study} of {Participatory} {Research} with {Disabled} {Students}}.
\newblock \bibinfo{journal}{\emph{Education Sciences}} \bibinfo{volume}{11}, \bibinfo{number}{10} (\bibinfo{date}{Oct.} \bibinfo{year}{2021}), \bibinfo{pages}{587}.
\newblock
\showISSN{2227-7102}
\urldef\tempurl%
\url{https://doi.org/10.3390/educsci11100587}
\showDOI{\tempurl}
\newblock
\shownote{Number: 10 Publisher: Multidisciplinary Digital Publishing Institute}.


\bibitem[Lockwood~Estrin et~al\mbox{.}(2021)]%
        {lockwood2021barriers}
\bibfield{author}{\bibinfo{person}{Georgia Lockwood~Estrin}, \bibinfo{person}{Victoria Milner}, \bibinfo{person}{Debbie Spain}, \bibinfo{person}{Francesca Happ{\'e}}, {and} \bibinfo{person}{Emma Colvert}.} \bibinfo{year}{2021}\natexlab{}.
\newblock \showarticletitle{Barriers to autism spectrum disorder diagnosis for young women and girls: A systematic review}.
\newblock \bibinfo{journal}{\emph{Review journal of autism and developmental disorders}} \bibinfo{volume}{8}, \bibinfo{number}{4} (\bibinfo{year}{2021}), \bibinfo{pages}{454--470}.
\newblock


\bibitem[Markelius(2024)]%
        {markelius_design_justice}
\bibfield{author}{\bibinfo{person}{Alva Markelius}.} \bibinfo{year}{2024}\natexlab{}.
\newblock \showarticletitle{An Empirical Design Justice Approach to Identifying Ethical Considerations in the Intersection of Large Language Models and Social Robotics}.
\newblock In \bibinfo{booktitle}{\emph{Oxford Intersections: AI in Society}}, \bibfield{editor}{\bibinfo{person}{Henry Shevlin}} (Ed.). \bibinfo{publisher}{Oxford University Press}, \bibinfo{address}{Oxford, England}.
\newblock


\bibitem[Markelius and Gunes(2025)]%
        {markscoping}
\bibfield{author}{\bibinfo{person}{Alva Markelius} {and} \bibinfo{person}{Hatice Gunes}.} \bibinfo{year}{2025}\natexlab{}.
\newblock \bibinfo{title}{Social Robotics and Large Language Models for Disability: A Scoping Review}.  (\bibinfo{year}{2025}).
\newblock
\newblock
\shownote{Fourthcoming}.


\bibitem[Matheus et~al\mbox{.}(2022)]%
        {Matheus_Vázquez_Scassellati_2022}
\bibfield{author}{\bibinfo{person}{Kayla Matheus}, \bibinfo{person}{Marynel Vázquez}, {and} \bibinfo{person}{Brian Scassellati}.} \bibinfo{year}{2022}\natexlab{}.
\newblock \showarticletitle{A Social Robot for Anxiety Reduction via Deep Breathing}. In \bibinfo{booktitle}{\emph{2022 31st IEEE International Conference on Robot and Human Interactive Communication (RO-MAN)}}. \bibinfo{publisher}{IEEE}, \bibinfo{address}{Napoli, Italy}, \bibinfo{pages}{89–94}.
\newblock
\showISSN{1944-9437}
\urldef\tempurl%
\url{https://doi.org/10.1109/RO-MAN53752.2022.9900638}
\showDOI{\tempurl}


\bibitem[McRuer(2008)]%
        {mcruer2008crip}
\bibfield{author}{\bibinfo{person}{Robert McRuer}.} \bibinfo{year}{2008}\natexlab{}.
\newblock \bibinfo{booktitle}{\emph{Crip theory. Cultural signs of queerness and disability}}.
\newblock \bibinfo{publisher}{Taylor \& Francis}, \bibinfo{address}{Oxfordshire, UK}.
\newblock


\bibitem[Meekosha and Shuttleworth(2009)]%
        {meekosha2009s}
\bibfield{author}{\bibinfo{person}{Helen Meekosha} {and} \bibinfo{person}{Russell Shuttleworth}.} \bibinfo{year}{2009}\natexlab{}.
\newblock \showarticletitle{What's so ‘critical’about critical disability studies?}
\newblock \bibinfo{journal}{\emph{Australian Journal of Human Rights}} \bibinfo{volume}{15}, \bibinfo{number}{1} (\bibinfo{year}{2009}), \bibinfo{pages}{47--75}.
\newblock


\bibitem[Mei et~al\mbox{.}(2023)]%
        {Mei_Fereidooni_Caliskan_2023}
\bibfield{author}{\bibinfo{person}{Katelyn~X. Mei}, \bibinfo{person}{Sonia Fereidooni}, {and} \bibinfo{person}{Aylin Caliskan}.} \bibinfo{year}{2023}\natexlab{}.
\newblock \showarticletitle{Bias Against 93 Stigmatized Groups in Masked Language Models and Downstream Sentiment Classification Tasks}. In \bibinfo{booktitle}{\emph{2023 ACM Conference on Fairness, Accountability, and Transparency}}. \bibinfo{publisher}{ACM}, \bibinfo{address}{Chicago, IL, USA}, \bibinfo{pages}{1699–1710}.
\newblock
\urldef\tempurl%
\url{https://doi.org/10.1145/3593013.3594109}
\showDOI{\tempurl}
\newblock
\shownote{arXiv:2306.05550 [cs]}.


\bibitem[Milton(2012)]%
        {Milton_2012}
\bibfield{author}{\bibinfo{person}{Damian~E.M. Milton}.} \bibinfo{year}{2012}\natexlab{}.
\newblock \showarticletitle{On the ontological status of autism: the ‘double empathy problem’}.
\newblock \bibinfo{journal}{\emph{Disability \& Society}} \bibinfo{volume}{27}, \bibinfo{number}{6} (\bibinfo{date}{Oct.} \bibinfo{year}{2012}), \bibinfo{pages}{883–887}.
\newblock
\showISSN{0968-7599, 1360-0508}
\urldef\tempurl%
\url{https://doi.org/10.1080/09687599.2012.710008}
\showDOI{\tempurl}


\bibitem[Mitchell et~al\mbox{.}(2021)]%
        {mitchell_social_2021}
\bibfield{author}{\bibinfo{person}{Alicia Mitchell}, \bibinfo{person}{Laurianne Sitbon}, \bibinfo{person}{Saminda~Sundeepa Balasuriya}, \bibinfo{person}{Stewart Koplick}, {and} \bibinfo{person}{Chris Beaumont}.} \bibinfo{year}{2021}\natexlab{}.
\newblock \showarticletitle{Social {Robots} in {Learning} {Experiences} of {Adults} with {Intellectual} {Disability}: {An} {Exploratory} {Study}}. In \bibinfo{booktitle}{\emph{Human-{Computer} {Interaction} – {INTERACT} 2021}}, \bibfield{editor}{\bibinfo{person}{Carmelo Ardito}, \bibinfo{person}{Rosa Lanzilotti}, \bibinfo{person}{Alessio Malizia}, \bibinfo{person}{Helen Petrie}, \bibinfo{person}{Antonio Piccinno}, \bibinfo{person}{Giuseppe Desolda}, {and} \bibinfo{person}{Kori Inkpen}} (Eds.). \bibinfo{publisher}{Springer International Publishing}, \bibinfo{address}{Cham}, \bibinfo{pages}{266--285}.
\newblock
\showISBNx{978-3-030-85623-6}
\urldef\tempurl%
\url{https://doi.org/10.1007/978-3-030-85623-6_17}
\showDOI{\tempurl}


\bibitem[Nakamura(2019)]%
        {Nakamura_2019}
\bibfield{author}{\bibinfo{person}{Karen Nakamura}.} \bibinfo{year}{2019}\natexlab{}.
\newblock \showarticletitle{My Algorithms Have Determined You’re Not Human: AI-ML, Reverse Turing-Tests, and the Disability Experience}. In \bibinfo{booktitle}{\emph{Proceedings of the 21st International ACM SIGACCESS Conference on Computers and Accessibility}} \emph{(\bibinfo{series}{ASSETS ’19})}. \bibinfo{publisher}{Association for Computing Machinery}, \bibinfo{address}{New York, NY, USA}, \bibinfo{pages}{1–2}.
\newblock
\showISBNx{978-1-4503-6676-2}
\urldef\tempurl%
\url{https://doi.org/10.1145/3308561.3353812}
\showDOI{\tempurl}


\bibitem[Neha et~al\mbox{.}(2024)]%
        {Neha_Kumar_Sankat_2024}
\bibfield{author}{\bibinfo{person}{Kandula Neha}, \bibinfo{person}{Ram Kumar}, {and} \bibinfo{person}{Monica Sankat}.} \bibinfo{year}{2024}\natexlab{}.
\newblock \showarticletitle{AI Wizards: Pioneering Assistive Technologies for Higher Education Inclusion of Students with Learning Disabilities}.
\newblock In \bibinfo{booktitle}{\emph{Applied Assistive Technologies and Informatics for Students with Disabilities}}, \bibfield{editor}{\bibinfo{person}{Rajesh Kaluri}, \bibinfo{person}{Mufti Mahmud}, \bibinfo{person}{Thippa~Reddy Gadekallu}, \bibinfo{person}{Dharmendra~Singh Rajput}, {and} \bibinfo{person}{Kuruva Lakshmanna}} (Eds.). \bibinfo{publisher}{Springer Nature}, \bibinfo{address}{Singapore}, \bibinfo{pages}{59–70}.
\newblock
\showISBNx{978-981-9709-14-4}
\urldef\tempurl%
\url{https://doi.org/10.1007/978-981-97-0914-4_4}
\showDOI{\tempurl}


\bibitem[Nordmann et~al\mbox{.}(2022)]%
        {nordmann2022lecture}
\bibfield{author}{\bibinfo{person}{Emily Nordmann}, \bibinfo{person}{Jacqui Hutchison}, {and} \bibinfo{person}{Jill~RD MacKay}.} \bibinfo{year}{2022}\natexlab{}.
\newblock \showarticletitle{Lecture rapture: The place and case for lectures in the new normal}.
\newblock \bibinfo{journal}{\emph{Teaching in Higher Education}} \bibinfo{volume}{27}, \bibinfo{number}{5} (\bibinfo{year}{2022}), \bibinfo{pages}{709--716}.
\newblock


\bibitem[Oliver(2013)]%
        {oliver2013social}
\bibfield{author}{\bibinfo{person}{Mike Oliver}.} \bibinfo{year}{2013}\natexlab{}.
\newblock \showarticletitle{The social model of disability: Thirty years on}.
\newblock \bibinfo{journal}{\emph{Disability \& society}} \bibinfo{volume}{28}, \bibinfo{number}{7} (\bibinfo{year}{2013}), \bibinfo{pages}{1024--1026}.
\newblock


\bibitem[Ostrowski et~al\mbox{.}(2022)]%
        {ostrowski2022ethics}
\bibfield{author}{\bibinfo{person}{Anastasia~K Ostrowski}, \bibinfo{person}{Raechel Walker}, \bibinfo{person}{Madhurima Das}, \bibinfo{person}{Maria Yang}, \bibinfo{person}{Cynthia Breazea}, \bibinfo{person}{Hae~Won Park}, {and} \bibinfo{person}{Aditi Verma}.} \bibinfo{year}{2022}\natexlab{}.
\newblock \showarticletitle{Ethics, equity, \& justice in human-robot interaction: A review and future directions}. In \bibinfo{booktitle}{\emph{2022 31st IEEE International Conference on Robot and Human Interactive Communication (RO-MAN)}}. \bibinfo{publisher}{IEEE}, \bibinfo{address}{Napoli, Italy}, \bibinfo{pages}{969--976}.
\newblock
\urldef\tempurl%
\url{https://doi.org/10.1109/RO-MAN53752.2022.9900805}
\showDOI{\tempurl}


\bibitem[O’Connell et~al\mbox{.}(2024)]%
        {O’Connell_Banga_Ayissi_Yaminrafie_Ko_Le_Cislowski_Mataric_2024}
\bibfield{author}{\bibinfo{person}{Amy O’Connell}, \bibinfo{person}{Ashveen Banga}, \bibinfo{person}{Jennifer Ayissi}, \bibinfo{person}{Nikki Yaminrafie}, \bibinfo{person}{Ellen Ko}, \bibinfo{person}{Andrew Le}, \bibinfo{person}{Bailey Cislowski}, {and} \bibinfo{person}{Maja Mataric}.} \bibinfo{year}{2024}\natexlab{}.
\newblock \showarticletitle{Design and Evaluation of a Socially Assistive Robot Schoolwork Companion for College Students with ADHD}. In \bibinfo{booktitle}{\emph{Proceedings of the 2024 ACM/IEEE International Conference on Human-Robot Interaction}}. \bibinfo{publisher}{ACM}, \bibinfo{address}{Boulder CO USA}, \bibinfo{pages}{533–541}.
\newblock
\showISBNx{9798400703225}
\urldef\tempurl%
\url{https://doi.org/10.1145/3610977.3634929}
\showDOI{\tempurl}


\bibitem[Padmanabha et~al\mbox{.}(2024)]%
        {Padmanabha_Yuan_Gupta_Karachiwalla_Majidi_Admoni_Erickson_2024}
\bibfield{author}{\bibinfo{person}{Akhil Padmanabha}, \bibinfo{person}{Jessie Yuan}, \bibinfo{person}{Janavi Gupta}, \bibinfo{person}{Zulekha Karachiwalla}, \bibinfo{person}{Carmel Majidi}, \bibinfo{person}{Henny Admoni}, {and} \bibinfo{person}{Zackory Erickson}.} \bibinfo{year}{2024}\natexlab{}.
\newblock \showarticletitle{VoicePilot: Harnessing LLMs as Speech Interfaces for Physically Assistive Robots}. In \bibinfo{booktitle}{\emph{Proceedings of the 37th Annual ACM Symposium on User Interface Software and Technology}} \emph{(\bibinfo{series}{UIST ’24})}. \bibinfo{publisher}{Association for Computing Machinery}, \bibinfo{address}{New York, NY, USA}, \bibinfo{pages}{1–18}.
\newblock
\showISBNx{9798400706288}
\urldef\tempurl%
\url{https://doi.org/10.1145/3654777.3676401}
\showDOI{\tempurl}


\bibitem[Rajagopal et~al\mbox{.}(2023)]%
        {A_V_Jebadurai_Vedamanickam_Kumar_2023}
\bibfield{author}{\bibinfo{person}{A Rajagopal}, \bibinfo{person}{V Nirmala}, \bibinfo{person}{Immanuel~Johnraja Jebadurai}, \bibinfo{person}{Arun~Muthuraj Vedamanickam}, {and} \bibinfo{person}{Prajakta~Uthaya Kumar}.} \bibinfo{year}{2023}\natexlab{}.
\newblock \showarticletitle{Design of Generative Multimodal AI Agents to Enable Persons with Learning Disability}. In \bibinfo{booktitle}{\emph{Companion Publication of the 25th International Conference on Multimodal Interaction}} \emph{(\bibinfo{series}{ICMI ’23 Companion})}. \bibinfo{publisher}{Association for Computing Machinery}, \bibinfo{address}{New York, NY, USA}, \bibinfo{pages}{259–271}.
\newblock
\showISBNx{9798400703218}
\urldef\tempurl%
\url{https://doi.org/10.1145/3610661.3617514}
\showDOI{\tempurl}


\bibitem[Rakova et~al\mbox{.}(2023)]%
        {rakova2023terms}
\bibfield{author}{\bibinfo{person}{Bogdana Rakova}, \bibinfo{person}{Renee Shelby}, {and} \bibinfo{person}{Megan Ma}.} \bibinfo{year}{2023}\natexlab{}.
\newblock \showarticletitle{Terms-we-serve-with: Five dimensions for anticipating and repairing algorithmic harm}.
\newblock \bibinfo{journal}{\emph{Big Data \& Society}} \bibinfo{volume}{10}, \bibinfo{number}{2} (\bibinfo{year}{2023}), \bibinfo{pages}{1--14}.
\newblock
\urldef\tempurl%
\url{https://doi.org/10.1177/20539517231211553}
\showDOI{\tempurl}


\bibitem[Rasouli et~al\mbox{.}(2024a)]%
        {Rasouli_Ghafurian_Nilsen_Dautenhahn_2024}
\bibfield{author}{\bibinfo{person}{Samira Rasouli}, \bibinfo{person}{Moojan Ghafurian}, \bibinfo{person}{Elizabeth~S. Nilsen}, {and} \bibinfo{person}{Kerstin Dautenhahn}.} \bibinfo{year}{2024}\natexlab{a}.
\newblock \showarticletitle{University Students’ Opinions on Using Intelligent Agents to Cope with Stress and Anxiety in Social Situations}.
\newblock \bibinfo{journal}{\emph{Computers in Human Behavior}}  \bibinfo{volume}{153} (\bibinfo{date}{April} \bibinfo{year}{2024}), \bibinfo{pages}{108072}.
\newblock
\showISSN{0747-5632}
\urldef\tempurl%
\url{https://doi.org/10.1016/j.chb.2023.108072}
\showDOI{\tempurl}


\bibitem[Rasouli et~al\mbox{.}(2024b)]%
        {rasouli2024university}
\bibfield{author}{\bibinfo{person}{Samira Rasouli}, \bibinfo{person}{Moojan Ghafurian}, \bibinfo{person}{Elizabeth~S Nilsen}, {and} \bibinfo{person}{Kerstin Dautenhahn}.} \bibinfo{year}{2024}\natexlab{b}.
\newblock \showarticletitle{University Students’ Opinions on Using Intelligent Agents to Cope with Stress and Anxiety in Social Situations}.
\newblock \bibinfo{journal}{\emph{Computers in Human Behavior}}  \bibinfo{volume}{153} (\bibinfo{year}{2024}), \bibinfo{pages}{108072}.
\newblock


\bibitem[Rizvi et~al\mbox{.}(2024)]%
        {rizvi2024robots}
\bibfield{author}{\bibinfo{person}{Naba Rizvi}, \bibinfo{person}{William Wu}, \bibinfo{person}{Mya Bolds}, \bibinfo{person}{Raunak Mondal}, \bibinfo{person}{Andrew Begel}, {and} \bibinfo{person}{Imani~NS Munyaka}.} \bibinfo{year}{2024}\natexlab{}.
\newblock \showarticletitle{Are Robots Ready to Deliver Autism Inclusion?: A Critical Review}. In \bibinfo{booktitle}{\emph{Proceedings of the CHI Conference on Human Factors in Computing Systems}}. \bibinfo{publisher}{ACM}, \bibinfo{address}{Honolulu, HI, USA}, \bibinfo{pages}{1--18}.
\newblock


\bibitem[S{\ae}tra et~al\mbox{.}(2022)]%
        {saetra2022normativity}
\bibfield{author}{\bibinfo{person}{Henrik~Skaug S{\ae}tra}, \bibinfo{person}{Anders Nordahl-Hansen}, \bibinfo{person}{Eduard Fosch-Villaronga}, {and} \bibinfo{person}{Christine Dahl}.} \bibinfo{year}{2022}\natexlab{}.
\newblock \showarticletitle{Normativity assumptions in the design and application of social robots for autistic children}. In \bibinfo{booktitle}{\emph{Proceedings of the 18th Scandinavian Conference on Health Informatics}}. \bibinfo{publisher}{Linköping Electronic Conference Proceedings 187}, \bibinfo{address}{Linkoping, Sweden}, \bibinfo{pages}{136--140}.
\newblock
\urldef\tempurl%
\url{https://doi.org/10.3384/ecp187023}
\showDOI{\tempurl}


\bibitem[Sasson and Morrison(2019)]%
        {sasson2019first}
\bibfield{author}{\bibinfo{person}{Noah~J Sasson} {and} \bibinfo{person}{Kerrianne~E Morrison}.} \bibinfo{year}{2019}\natexlab{}.
\newblock \showarticletitle{First impressions of adults with autism improve with diagnostic disclosure and increased autism knowledge of peers}.
\newblock \bibinfo{journal}{\emph{Autism}} \bibinfo{volume}{23}, \bibinfo{number}{1} (\bibinfo{year}{2019}), \bibinfo{pages}{50--59}.
\newblock


\bibitem[Shaw(2024)]%
        {shaw2024inclusion}
\bibfield{author}{\bibinfo{person}{Anne Shaw}.} \bibinfo{year}{2024}\natexlab{}.
\newblock \showarticletitle{Inclusion of disabled Higher Education students: why are we not there yet?}
\newblock \bibinfo{journal}{\emph{International Journal of Inclusive Education}} \bibinfo{volume}{28}, \bibinfo{number}{6} (\bibinfo{year}{2024}), \bibinfo{pages}{820--838}.
\newblock
\urldef\tempurl%
\url{https://doi.org/10.1080/13603116.2021.1968514}
\showDOI{\tempurl}


\bibitem[Sheppard et~al\mbox{.}(2016)]%
        {sheppard2016easy}
\bibfield{author}{\bibinfo{person}{Elizabeth Sheppard}, \bibinfo{person}{Dhanya Pillai}, \bibinfo{person}{Genevieve Tze-Lynn Wong}, \bibinfo{person}{Danielle Ropar}, {and} \bibinfo{person}{Peter Mitchell}.} \bibinfo{year}{2016}\natexlab{}.
\newblock \showarticletitle{How easy is it to read the minds of people with autism spectrum disorder?}
\newblock \bibinfo{journal}{\emph{Journal of autism and developmental disorders}}  \bibinfo{volume}{46} (\bibinfo{year}{2016}), \bibinfo{pages}{1247--1254}.
\newblock


\bibitem[Sloane et~al\mbox{.}(2022)]%
        {sloane2022participation}
\bibfield{author}{\bibinfo{person}{Mona Sloane}, \bibinfo{person}{Emanuel Moss}, \bibinfo{person}{Olaitan Awomolo}, {and} \bibinfo{person}{Laura Forlano}.} \bibinfo{year}{2022}\natexlab{}.
\newblock \showarticletitle{Participation is not a design fix for machine learning}. In \bibinfo{booktitle}{\emph{Proceedings of the 2nd ACM Conference on Equity and Access in Algorithms, Mechanisms, and Optimization}}. \bibinfo{pages}{1--6}.
\newblock


\bibitem[Smith and Mueller(2022)]%
        {smith2022importance}
\bibfield{author}{\bibinfo{person}{Ivanova Smith} {and} \bibinfo{person}{Carlyn~O Mueller}.} \bibinfo{year}{2022}\natexlab{}.
\newblock \showarticletitle{The importance of disability identity, self-advocacy, and disability activism}.
\newblock \bibinfo{journal}{\emph{Inclusive Practices}} \bibinfo{volume}{1}, \bibinfo{number}{2} (\bibinfo{year}{2022}), \bibinfo{pages}{47--54}.
\newblock


\bibitem[Spitale et~al\mbox{.}(2023)]%
        {spitale2023vita}
\bibfield{author}{\bibinfo{person}{Micol Spitale}, \bibinfo{person}{Minja Axelsson}, {and} \bibinfo{person}{Hatice Gunes}.} \bibinfo{year}{2023}\natexlab{}.
\newblock \bibinfo{title}{Vita: A multi-modal llm-based system for longitudinal, autonomous, and adaptive robotic mental well-being coaching}.  (\bibinfo{year}{2023}).
\newblock
\newblock
\shownote{arXiv preprint arXiv:2312.09740}.


\bibitem[Takeuchi et~al\mbox{.}(2020)]%
        {Takeuchi_Yamazaki_Yoshifuji_2020}
\bibfield{author}{\bibinfo{person}{Kazuaki Takeuchi}, \bibinfo{person}{Yoichi Yamazaki}, {and} \bibinfo{person}{Kentaro Yoshifuji}.} \bibinfo{year}{2020}\natexlab{}.
\newblock \showarticletitle{Avatar Work: Telework for Disabled People Unable to Go Outside by Using Avatar Robots}. In \bibinfo{booktitle}{\emph{Companion of the 2020 ACM/IEEE International Conference on Human-Robot Interaction}}. \bibinfo{publisher}{ACM}, \bibinfo{address}{Cambridge United Kingdom}, \bibinfo{pages}{53–60}.
\newblock
\showISBNx{978-1-4503-7057-8}
\urldef\tempurl%
\url{https://doi.org/10.1145/3371382.3380737}
\showDOI{\tempurl}


\bibitem[Test et~al\mbox{.}(2005)]%
        {test2005conceptual}
\bibfield{author}{\bibinfo{person}{David~W Test}, \bibinfo{person}{Catherine~H Fowler}, \bibinfo{person}{Wendy~M Wood}, \bibinfo{person}{Denise~M Brewer}, {and} \bibinfo{person}{Steven Eddy}.} \bibinfo{year}{2005}\natexlab{}.
\newblock \showarticletitle{A conceptual framework of self-advocacy for students with disabilities}.
\newblock \bibinfo{journal}{\emph{Remedial and Special education}} \bibinfo{volume}{26}, \bibinfo{number}{1} (\bibinfo{year}{2005}), \bibinfo{pages}{43--54}.
\newblock


\bibitem[{The University of Cambridge’s Disability Service}(2010)]%
        {DRC2010}
\bibfield{author}{\bibinfo{person}{{The University of Cambridge’s Disability Service}}.} \bibinfo{year}{2010}\natexlab{}.
\newblock \bibinfo{title}{DRC 2009/10 Annual Report}.
\newblock
\newblock
\urldef\tempurl%
\url{https://www.disability.admin.cam.ac.uk/sites/default/files/2023-09/adrc-annual-report-2009-2010-pdf-283kb.pdf}
\showURL{%
\tempurl}
\newblock
\shownote{Accessed: 2025-02-18}.


\bibitem[Urbina et~al\mbox{.}(2024)]%
        {Urbina_Vu_Nguyen_2024}
\bibfield{author}{\bibinfo{person}{Jacob~T. Urbina}, \bibinfo{person}{Peter~D. Vu}, {and} \bibinfo{person}{Michael~V. Nguyen}.} \bibinfo{year}{2024}\natexlab{}.
\newblock \showarticletitle{Disability Ethics and Education in the Age of Artificial Intelligence: Identifying Ability Bias in ChatGPT and Gemini}.
\newblock \bibinfo{journal}{\emph{Archives of Physical Medicine and Rehabilitation}} \bibinfo{volume}{106}, \bibinfo{number}{1} (\bibinfo{year}{2024}), \bibinfo{pages}{14--19}.
\newblock
\showISSN{0003-9993}
\urldef\tempurl%
\url{https://doi.org/10.1016/j.apmr.2024.08.014}
\showDOI{\tempurl}


\bibitem[Williams(2021)]%
        {williams2021misfit}
\bibfield{author}{\bibinfo{person}{Rua~M Williams}.} \bibinfo{year}{2021}\natexlab{}.
\newblock \bibinfo{title}{I, Misfit: Empty Fortresses, Social Robots, and Peculiar Relations in Autism Research.}
\newblock
\newblock


\end{thebibliography}

\appendix

\section{Research Methods}

\subsection{Glossary of Terms}
This glossary of terms was provided both for the interviews and both focus group for all participants to have an understanding of the most important concepts covered in the research.

\begin{itemize}
    \item \textbf{Support needs} refers to any aspect of study and life at university that might present a challenge for disabled students, including accessing information about the university, college, department or course, accessing additional services or resources targeted at disabled students, specific advice on strategies or adjustments, assistance in contacting or communicating with department, college or university staff, etc
    \item \textbf{Self-advocacy} includes any communication (written, verbal or otherwise) involving a student expressing their learning needs or preferences in order to better access their learning.
    \item \textbf{Mediation} refers to resolving an issue, tension or conflict with a department or college through communication (either verbal or text, e.g. email).
    \item \textbf{AI-based tools} refers to any software that uses algorithms, data, and computational power to simulate human intelligence, e.g. ChatGPT.
    \item \textbf{‘Other technologies’} include assistive technologies such as dictation or audio software, and any other software or equipment that might support access to learning for disabled students, e.g. Dragon speech recognition software.
    \item \textbf{Social robots} are machines that are aimed at interacting socially with human users, they can (but do not necessarily) communicate with high-level dialogue, establish/maintain social relationships, use non-verbal cues, express and perceive emotions and exhibit distinctive personality and character.
    \item \textbf{Large Language Models} are generative AI-systems which use deep learning and large amounts of data to process and generate natural language and other types of content (e.g. visual) to perform a wide range of tasks, such as communication and reasoning.

\end{itemize}

\subsection{Interview Prompts}

\subsubsection{Demographics }

Gender identity; age; current role/s; qualifications and previous experience with respect to disabled students

(a) Before we get to the questions, can I ask your age and how you define your gender?

(b) What are you comfortable to share about your disability? Are you happy to indicate which category/ies of disability apply/ies to you 
\subsubsection{Advocacy and communication needs}

What are the most significant barriers to effective communication and self-advocacy for disabled students?
We’ll start by talking about advocacy and communication

(a) Can you tell me about a time when advocacy and communication about your support needs went really well? Who was involved? What did they do? How did you feel?

(b) can you tell me about a time when advocacy and communication about your support needs did not go well? Who was involved? What did they do? How did you feel?

(c) Have you found any technology (software, apps, equipment) helpful in assisting your learning or communicating your needs?

(d) In an ideal world, what issues with communication would a tech-based innovation solve? what features would it have? What would be particularly helpful for you?
\subsubsection{The use of AI-based tools}

What are the benefits and challenges of current forms of AI used by disabled students to support communication?
Next, focusing on AI tools

(a) Have you used any AI tools, such as ChatGPT or Goblin Tools? Which ones? 

(b) Have you used any of these tools to support communication e.g. writing emails? Was the tool helpful? In what way?

(c) Were there any barriers to using this tool? Or any difficulties? What were those?

(d) Did you have any worries about using the tool?

(e) Do you have any worries about making more use of AI tools for supporting communicating about your study needs?
\subsubsection{The use of social robots}

What are the perceptions and concerns around the use of social robots with disabled students? 
I’m going to show you a short clip of social robots in action

(a) Have you ever seen or interacted with a social robot -  that is, one similar to those we saw in the video? Can you tell me about what it was like? How did you feel about interacting with a robot? If not, what do you think it would be like?

(b) Are you aware that social robots can be used to support communication and provide coaching? Would you feel comfortable asking a robot for advice or as a sounding board?

(c) If you were interacting with a social robot, how would you feel about the robot understanding or responding to your verbal (e.g., yes/no, i agree etc.) and nonverbal behaviours   (e.g., sigh, frown or laughter etc.)? For example, should it recognize when you need a break or encouragement or when you are upset?

(d) Would you have any concerns about interacting with a social robot similar to the ones you saw in the video?
\subsubsection{The use of robots in student support}

What role might social robots have in supporting disabled students?
Thinking about how robots might be useful in the future

(a) What do you think might be useful about having a social robot around your college or department to assist you with communicating your needs and accessing adjustments?

(b) Do you think a robot could do any of the tasks that are currently done by the person who supports you in accessing learning at university? What tasks could it do?

(c) Do you think a robot would be better than a person for any aspect of your learning support?  How about helping you with communication or self-advocacy?
\subsubsection{Ethical and practical concerns around robots and AI}

What are the practical and ethical concerns around social robots and AI?
Finally, I would like to ask you about practical and ethical concerns you may have.

(a) Do you think having a social robot available would make it more convenient or easier to get help? What sorts of help do you think it might be useful for?

(b) Do you think we should be using robots for roles that are currently done by people?  Which role/s? Why/why not?

(c) What would you worry about if there was a social robot around your college or department? What would you want to avoid?

(d) What features, limits or safeguards would need to be in place for you to feel comfortable interacting with a social robot?

NOTE: questions were adapted for student support professionals,

\subsection{Focus Group 1 Topic Guide}

\begin{longtable}{|p{0.15\textwidth}|p{0.40\textwidth}|p{0.45\textwidth}|}
\hline
\textbf{Section} & \textbf{Key questions} & \textbf{Notes} \\
\hline
\endfirsthead
\hline
\textbf{Section} & \textbf{Key questions} & \textbf{Notes} \\
\hline
\endhead
\hline
\endfoot

\textbf{Introductions and context:} General perceptions of technology, AI, robotics &  - How familiar are you with technologies currently or potentially being used in educational settings for students with disabilities? \newline
- What specific challenges could technology help address for students with disabilities? \newline
- Do you think technology can be empowering or limiting for students with disabilities? Why? \newline & 
Welcome and housekeeping. \newline
- Interactive activity: Mentimeter, indicate what technologies you have found useful previously: examples, social robot, Goblin Tools, Notion, ChatGPT, Dragon Speech \newline
- Use images again to support open questioning, e.g.  'which of these technologies would you use? Why/why not? \newline
- Follow up probing questions to get a sense of areas of consensus/divide
 \\
\hline

\textbf{Focus area 1} Understanding the needs and problem space & - What kind of mediation and advocacy is needed for students with disabilities? \newline
- What do students think are the most important factors to consider when supporting students with disabilities through mediation and advocacy?  & 
- Invite students to think about a time when advocacy or mediation was needed to enable them to successfully access their studies. \newline
- Invite participants to think about a time when they were able to successfully ask for adjustments, support or adaptations and consider what made it successful. \newline
- Encourage interaction to probe commonalities or differences in experience and perspectives.\newline
- Optional: Invite sharing of challenges where participants feel comfortable in doing so in this group. Invite discussion on communication of needs to relevant staff, barriers and facilitators and experiences. 
 \\
\hline

\textbf{Focus area 2:} Ideating robotic support for disabled students & - What are the most important considerations of using technologies for disabled students' advocacy and mediation? \newline
- What does success look like when using a robot in advocacy or mediation? \newline 
- What specific outcomes would you hope to achieve?  & - Show video clip of social robots in action. \newline
- Guide participants in discussion of their reactions to the robots and a potential role in student support; \newline
- What are their initial reactions to the robot? \newline
- How do they feel about the idea of working with one? \newline
- What do they think about the potential role of a robot in supporting them? (e.g., informative - provides information; supportive - coaches them in learning and/or practising certain skills; facilitation - act as intermediary in interactions; other - provides space for letting steam off; etc.) \newline
- What would need to be in place for this (environment, safeguards, etc.)
 \\
\hline

\textbf{Focus area 3} Features, limitations and capabilities& - What capabilities would you like to see in a robot for mediation and advocacy? How should it communicate and offer support? \newline
- Do you think there are specific ways a robot could better support communication, either with other students or teachers? & 
- Virtual post-it’s with capabilities (from interviews) and possibility to choose which ones are most important or adding new. \newline
- Invite discussion about features and capabilities: \newline
- Do you think you would be more comfortable talking to a robot about concerns you might not want to share with human representatives of the uni? Why and what would you if so want to share? \newline
- What kinds of things would a robot need to do or say to make you feel like it really understands you? \newline
- How important is it for you to feel in control during interactions with robots? What features would help you maintain that sense of control? \newline
- How do you feel about relying on a robot to assist you in mediation or advocacy? What would make you trust in this technology? \newline
- To what extent would you like to be able to customise a robot tool? Should it have a feature to be able to adjust/tailor its usage and capabilities to your specific needs and preferences? \\
\hline

\textbf{Focus area 4:} Design \& Potential Scenarios& - What do you want from the robot? Where should it be located? What topics should it cover? \newline
- Envision scenarios like signposting, expressing needs, and empathizing with shared experiences. & 
Present different potential scenarios to the focus group, allow the participants to envisage: \newline
- The robot does signposting, communication of information. \newline
- Double empathy problem: envision how you would overcome it with the help of a robot.\newline
- Rehearsing to help express your needs to relevant other people (staff, students, professors). \newline
- The robot has similar challenges and/or experiences as you, and you can use it to vent, let off steam or to share and empathise. \newline
- Do you have any thoughts or preferences regarding the robot's voice and appearance? How should it look like / sound like to not be distracting but rather be helpful and useful? \\
\hline

\textbf{Focus area 5:} Ethical considerations and limitations & - What limitations or concerns come to mind regarding the use of AI or robotic technologies in student advocacy, mediation and communication?  \newline
- How can these concerns be addressed to make you feel more at ease? \newline - What are considered the key benefits and risks/concerns? & 
Virtual post-it’s with potential ethical concerns (from interviews) and possibility to choose which ones are most important to them (privacy, bias, nuance, customisation, transparency, unfamiliarity, lack of trust, lack of relatability) Invite discussion to limitations:\newline
- What kind of information would make you feel comfortable using a robot or sharing information with it? Is there anything specific you would want to know beforehand? \newline
- How would you prefer to receive information about how a robot works and what data it collects? What would make this information clear and accessible to you? \newline
- Data usage and privacy, is there a tradeoff between personalisation/adaption and privacy? \newline
- How would you prefer to receive information about how a robot works and what data it collects? What would make this information clear and accessible to you? \newline
- Have you heard about biases in AI / robots and would you be concerned about this if using it as a tool?
 \\
\hline

\textbf{Wrap up} & Closing questions & - From our discussion today, what stands out as the most helpful use of technology for you? \newline
- What do you see as the biggest potential risks and benefits using robots for students with disabilities? \newline
- Is there anything else that you would like to add or any point that you feel needs further discussion? \\
\hline

\end{longtable}

\subsection{Focus Group 2 Topic Guide}

\begin{longtable}{|p{0.15\textwidth}|p{0.40\textwidth}|p{0.45\textwidth}|}
\hline
\textbf{Section} & \textbf{Key questions} & \textbf{Notes} \\
\hline
\endfirsthead
\hline
\textbf{Section} & \textbf{Key questions} & \textbf{Notes} \\
\hline
\endhead
\hline
\endfoot

\textbf{Introductions and context:} Short recap &  & - Welcome and housekeeping.\newline
- Short recap of previous focus group, invite discussion if anything is unclear or anyone wants to add anything. \newline
- Introduction to co-design, explaining the purpose of this focus group
 \\
\hline

\textbf{Robot demo} &  & A demo of the QT robot interacting with one of the experiments. The robot introduces itself as well as speaks about its potential capabilities.
 \\
\hline

\textbf{Focus area 1:} Robot role design dimensions
 &- What do you want from the robot in relation to your disability? \newline - What do you need it for in that context? \newline What topics would you want to talk about? \newline - What would the role of the robot be? \newline - How would you \textbf{envision} a social robot (based on the above questions)?  & 
Warm-up exercise with print-outs. Present 2 categories of design to the focus group (task/role, form), allow the participants to use print outs to make combinations of: \newline
- Roles: i) Signposting, ii) Advocacy practice, iii) Social communication practice, iv) Presentation rehearsal partner, v) Study Companion (plays music etc), vi) Venting partner \newline
- Form: QT, Nao, Furhat, Pepper
 \\
\hline

\textbf{Focus area 2:} Robot characteristics design dimensions & This focus area relates to design and envisaging of \textbf{physical and non-physical characteristics} of the robot itself. 
& 
- How do you envisage the robot form? Both based on the pictures provided but also if you could imagine freely? \newline
- Do you have any thoughts or preferences regarding the robot's voice and appearance? How should it look like / sound like to not be distracting but rather be helpful and useful? \newline
- How expressive should the robot be? Should it exhibit a lot of emotion through facial expressions and body language? \newline
- What languages should the robot be able to speak? Are there specific aspects of language that you find important e.g. code-switching, dialects, accents, slang, jargon etc? \newline
- What are your thoughts on the robot’s gender? Should it be gendered at all? \newline
- What characteristics or identity factors are important to relate to the robot? E.g. Does the robot have a gender, a disability, a race, a religion and if so, how is this expressed? \newline
 \\
\hline

\textbf{Focus area 3:} Robot contextual design dimensions & 
This focus area relates to design and envisaging of \textbf{contextual, situational and structural} factors of the robot. 
& 
- How do you envisage your relationship with the robot? E.g. is it more like a pet, a coach, a friend, an advocate, a computer, a tool? \newline
- Where should it be located? How do you envision the environment the robot is situated within? \newline
- How much control should students, researchers, or other stakeholders (e.g. the ADRC) have over the robot? \newline
- How can students maintain a feeling of control over the robot system? \newline
- What does it mean to agree/consent to use the robot in a way that feels meaningful to you? \newline
- If there would be a user agreement for using the robot, what would you like it to include? \newline
- How could terms and conditions be made clearer and more helpful when it comes to explaining how a social robot might impact people?\newline
- Who would be using the robot? Would you use it alone, or in the company of other people, e.g. friends, disability reps, ADRC personnel?\newline
\\
\hline

\textbf{Focus area 4:} Ethical considerations and limitations & This focus area concerns \textbf{limitations and ethical considerations} related to using a social robot for students with disabilities.
& 
- Are there any problems/ethical concerns/issues you might encounter when using the robot, like things not working right or unfairnesses? How do you think we could fix these problems? Who should be responsible if that happens? \newline
- What factors would influence your trust in the robot? What mechanisms would have to be in place for you to trust the robot? \newline
W- hat data should the robot collect and not collect? Personal information, video, audio etc? \newline
- What mechanisms would have to be in place for the robot to interact with disabled people and overcome communication assumptions?\newline
- If you had a problem or concern with using the social robot, e.g. related to racial or ableist bias, what would make it easy for you to share your concerns? \newline
- Can you think of any small reminders or “nudges” that have helped you understand or make better decisions when using AI tools? What kinds of features like this would help people feel more informed or in control of using the robot?\newline
- If the social robot system causes harm, what do you think should be done to make things right for the people affected?
 \\
\hline

\textbf{Wrap up} & Closing questions &  Is there anything else that you would like to add or any point that you feel needs further discussion? \\
\hline

\end{longtable}
Please note that focus group 2 was an open discussion and that all focus areas in the topic guide were not covered. The guide was used as a way to probe questions and topics, and not as a strict script to follow.

\end{document}